\def\mearth{{\rm\,M_\oplus}}
\def\mjup{{\rm\,M_{jup}}}
\def\msat{{\rm\,M_{saturn}}}
\def\bbc{{\beta / \beta_\mathrm{crit}}}
\def\gsim{~\rlap{$>$}{\lower 1.0ex\hbox{$\sim$}}}
\def\lsim{~\rlap{$<$}{\lower 1.0ex\hbox{$\sim$}}}

\documentclass[12pt, preprint]{aastex}
\usepackage{graphicx}
\usepackage{natbib,color,lscape}
\usepackage{subfigure}
\usepackage{threeparttable} 
\usepackage{url}

\begin{document}
\title{Stability analysis of single planet systems and their habitable zones}
\author{Ravi kumar Kopparapu\altaffilmark{1,2}, Rory Barnes\altaffilmark{2,3}}
\altaffiltext{1}{Department of Physics, 104 Davey lab, Pennsylvania State
University, University Park, PA - 16802-6300, USA; ravi@gravity.psu.edu}
\altaffiltext{2}{Virtual Planetary Laboratory}
\altaffiltext{3}{Department of Astronomy, University of Washington, Seattle,
WA, 98195-1580}

\begin{abstract}

 We study the dynamical stability of  planetary systems consisting of
 one  hypothetical
   terrestrial mass planet ($1~  $ or $10 \mearth$) and one 
massive planet ($10 \mearth - 10 \mjup$).   
 We consider masses and orbits that cover the range of observed
   planetary system architectures (including non-zero
initial eccentricities), determine the stability limit 
through N-body
   simulations, and compare it to the analytic Hill stability boundary.
   We show that for given masses and orbits of a two planet system,
  a single parameter, which can be calculated analytically,  describes the 
Lagrange stability boundary (no ejections or exchanges) but which
 diverges significantly from the Hill stability boundary.
However, we do find that the actual boundary is 
fractal, and therefore we also identify a second parameter which demarcates the transition
 from stable to unstable evolution.
We show the portions of the habitable zones of $\rho$ CrB, HD 164922, GJ 674, and HD 7924 which can
 support a terrestrial planet.
  These analyses clarify the stability boundaries in exoplanetary systems and
 demonstrate that, for most exoplanetary systems, numerical simulations of the
stability of potentially habitable planets are only necessary over a narrow region
of parameter space.
Finally we also identify and provide a catalog of known systems 
which can host terrestrial planets in their habitable zones.
\end{abstract}

\maketitle

\section{Introduction}
\label{sec1}

The dynamical stability of extra-solar planetary systems can constrain
planet formation models, reveal commonalities among planetary systems
and may even be used to infer the existence of unseen companions.
Many authors have studied the dynamical  stability 
of our solar system and extra-solar planetary systems 
 \citep[see][for example]{Wisdom1982,Laskar1989, RasioFord1996, Chambers1996,
 LaughlinChambers2001, Gozdziewski2001, Ji2002,BQ04,
Ford2005, Jones2006,raymond09}. These investigations have revealed that 
planetary systems
are close to dynamical instability, illuminated the boundaries
between stable and unstable configurations, and identified the parameter
space that can support additional planets.

From an astrobiological point of view, dynamically stable
 habitable zones  (HZs)  for 
terrestrial mass planets ($0.3 \mearth < M_{p} < 10 \mearth$) are the most interesting.
Classically, the HZ is defined as the circumstellar  region in which
 a terrestrial mass  planet with favorable atmospheric conditions
 can sustain liquid water on its surface \citep[][but see also 
Barnes et al.(2009)]{Huang1959, Hart1978, Kasting1993, Selsis2007}.

Previous work  \citep{Jones2001, MT2003, Jones2006, Sandor2007} investigated
the orbital stability of 
Earth-mass planets in the HZ  of systems with a Jupiter-mass companion.
In their pioneering work, \cite{Jones2001} 
estimated the stability of four known planetary systems in the HZ of their
host stars.
\cite{MT2003} considered the dynamical stability of 
100 terrestrial mass-planets (modelled as test particles) in the HZs of
 the then-known 85 extra-solar planetary systems. 
 From their simulations, they generated a tabular list
 of stable HZs for all observed systems.
However, that study did not systematically consider eccentricity, is not 
generalizable to arbitrary planet masses, and relies on numerical
experiments to determine stability.
A similar study by \cite{Jones2006} also examined 
the stability of  Earth-mass planets in the HZ.
 Their results indicated that $41 \%$
of the systems in their sample had ``sustained habitability''.
Their simulations were also not generalizable and based on a large set
of numerical experiments which assumed the potentially habitable planet
was on a circular orbit.
Most recently, \cite{Sandor2007} considered systems consisting of a giant planet 
 with a maximum 
eccentricity of $0.5$ and a terrestrial planet (modelled as a test particle 
initially in circular orbit)
They used relative Lyapunov indicators  and fast
Lyapunov indicators  to identify stable zones and generated a stability 
catalog, which can be applied to systems with mass-ratios in the range
$10^{-4} - 10^{-2}$ between the giant planet and the star.
Although this catalog is generalizable to massive planets between a Saturn-mass
and $10 \mjup$, it still assumes the terrestrial planet is on a circular orbit.

These studies made great strides toward a universal definition of HZ
stability. However, several aspects of each study could be improved, such as
a systematic assessment of the stability of terrestrial planets on eccentric
orbits, a method that eliminates the need for computationally expensive 
numerical experiments, and wide coverage of planetary masses. In this
investigation we address each of these points and develop a simple analytic
approach that applies to arbitrary configurations of a giant-plus-terrestrial 
planetary system.

As of March 2010, 376 extra-solar planetary systems have been detected,
and the majority (331, $\approx 88 \%$) are single planet systems.
 This opens up the possibility that there may be additional planets
not yet detected, in the stable regions of these systems.
According to \cite{wright2007} more than $30 \%$ of known single  
planet systems show evidence for additional companions. Furthermore, \cite{marcy2005} showed
that the distribution of observed planets rises steeply towards smaller masses.
The analyses of \cite{wright2007} \& \cite{marcy2005}
 suggests that many systems
may have low mass planets\footnote{ \cite{Wittenmyer2009} did a comprehensive study of 22
 planetary systems using the Hobby-Eberly 
Telescope (HET; \cite{Ramsey98}) and found no additional planets, 
 but their study had a  radial velocity (RV) precision of just
$10 \sim 20$ ms$^{-1}$, which can only detect low-mass planets in tight 
orbits.}.
Therefore, maps of stable regions in known planetary systems can aid
observers in their quest to discover more planets in known systems.

We consider two definitions of dynamical stability:
(1) {\it Hill stability}: A system is 
 Hill-stable if the ordering of planets
is conserved,
even if the outer-most planet  escapes to infinity. 
(2) {\it Lagrange stability}: In this kind of stability, every planet's
motion is bounded, i.e, no planet escapes from the system and exchanges are
forbidden.
Hill stability for a two-planet, non-resonant system can 
be described by an analytical expression \citep{MB1982, Gladman1993}, whereas
no analytical criteria are available for Lagrange stability so
 we investigate it through numerical simulations.
Previous studies by \cite{BG06, BG07}  showed that
Hill stability is a reasonable approximation to Lagrange stability 
in the case of two approximately Jupiter-mass planets.
 Part of the goal of our present work is to broaden
the parameter space considered by \cite{BG06, BG07}.

In this investigation,  we explore the stability of hypothetical 
$1 \mearth$ and $10  \mearth$ planets in the HZ and in the presence
of giant and super-Earth planets.  We consider nonzero initial eccentricities
of terrestrial planets and
find that a modified version of the Hill stability criterion 
  adequately describes the Lagrange stability boundary.
Furthermore, we provide an analytical expression that identifies
 the Lagrange stability boundary of two-planet, non-resonant systems. 

 Utilizing these  boundaries, we  provide a catalog of  fractions of
HZs that are Lagrange stable for terrestrial mass planets in all the currently 
known 
single planet systems.
This catalog can help guide observers toward systems that can host 
terrestrial-size planets in their HZ.

The plan of our paper is as follows: In Section \ref{sec2},
 we  discuss the Hill and Lagrange stability criteria,
 describe our numerical methods, and present our  model of the HZ.
 In Section \ref{sec3}, we present our results
 and explain how to identify  the Lagrange stability boundary for any system
with one $\ge 10 \mearth$ planet and one $\le 10 \mearth$ planet. 
In Section \ref{sec4}, we apply our results to some of the 
known single planet systems.
Finally,  in Section \ref{sec5}, we summarize the investigation, discuss its
importance  for observational programs, and 
suggest directions for future research.

\section{Methodology}
\label{sec2}

 According to \cite{MB1982}, a system is Hill stable if the following inequality is satisfied:

\begin{eqnarray}
- \frac{2 M}{G^{2} M^{3}_{\star}} ~  c^{2} h > 1 + 3^{4/3} \frac{m_{1} m_{2}}{m^{2/3}_{3}
 (m_{1} + m_{2})^{4/3}} + . . .
\label{bbc}
\end{eqnarray}
where $M$ is the total mass of the system, $G$ is the gravitational constant, $M_{\star} = 
m_{1} m_{2} + m_{2} m_{3} + m_{3} m_{1}$, $c$ is the total angular momentum of the system, $h$  is the
total energy, $m_{1}$, $m_{2}$ and $m_{3}$ are the masses of the planets and 
the star, respectively.
 We call the left hand side of 
Eq.(\ref{bbc}) $\beta$
 and the right hand-side  $\beta_{crit}$.
If $\bbc > 1$, then a system is definitely Hill stable, if not the Hill
stability is unknown.


Studies by \cite{BG06, BG07} found that for two Jupiter mass planets, 
if $\bbc \gsim 1$ 
(and no resonances are present), then the
 system is Lagrange stable. Moreover, Barnes et al. (2008a) found that systems tend to be
packed if $\bbc \lesssim 1.5$ and not packed when
$\bbc \gtrsim 2$. \cite{BG07} pointed out that the vast majority of
two-planet systems are observed with $\bbc < 1.5$ and hence are packed.
Recently, \cite{KRB09} proposed that the HD 47186 planetary system, with
 $\bbc = 6.13$ the largest value among known, non-controversial
systems that have not been affected by tides\footnote{See
\url{http://xsp.astro.washington.edu}
for an up to date list of  
$\bbc$ values for the known extra-solar multiple
planet systems.}, may have at least one additional (terrestrial mass)
companion  in the HZ between the two known planets.

To determine the dynamically stable regions around single planet systems,
 we numerically explore the 
mass, semi-major axis and
 eccentricity  space of  model systems, which  cover
the range of observed of  extra-solar planets.
In all the models (listed in Table \ref{table1}), we assume that the hypothetical additional planet is either 
$1 \mearth$ or $10 \mearth$  and consider the following massive companions, 
(which we presume are already known to exist):
(1) $10 \mjup$, (2) $5.6 \mjup$ (3) $3 \mjup$, (4) $1.77 \mjup$ 
(5) $1 \mjup$, (6) $1.86 \msat$ (7) $1 \msat$, (8) $56 \mearth$ 
(9) $30 \mearth$ (10) $17.7 \mearth$
and (11) $10 \mearth$.
Most simulations assume that the host star has the same 
mass, effective temperature ($T_\mathrm{eff}$) and luminosity as the Sun.
Orbital elements such as the  longitude of periastron $\varpi$ are chosen 
randomly before the beginning of the simulation (Eq. (1) only depends weakly
on them). 
For ``known'' Saturns  and super-Earths, we fix semi-major axis  $a$  
 at $0.5$ AU (and the HZ is exterior) or at $2$ AU (the HZ
interior). For super-Jupiter and Jupiter mass, $a$ is fixed either at $0.25$ AU or at 
$4$ AU. These choices allow  at least part of the HZ to be Lagrange stable.
 Although we 
choose configurations that focus on the HZ, the results should apply to all 
regions in the system.

We explore dynamical stability by performing a large number of N-body simulations,
each with a different initial condition. For the known planet, we keep $a$
 constant and vary its initial
 eccentricity, $e$, from $0 - 0.6$ in steps of $0.05$.
 We calculate $\bbc$  from
Eq.(\ref{bbc}), by varying the hypothetical planet's semi-major axis and  initial
eccentricity. 
 In order to find the Lagrange stability boundary,
 we perform numerical simulations along a particular $\bbc$ curve, with {\tt Mercury}
\citep{Chambers1999}, using the hybrid integrator.
 We integrate each configuration for $10^{7}$ years, long enough
 to identify unstable regions \citep{BQ04}.
The time step was  small enough that energy is 
 conserved to better than 1 part in $10^{6}$.
 A system is considered Lagrange unstable if the semi-major axis of the
terrestrial mass planet changes by 15\% of the initial value or if the two 
planets come within $3.5$ Hill radii of each other\footnote{A recent study by \cite{CY2009} notes that the Hill-radius criterion for ejection of a Earth
mass planet around a giant planet may not be valid. Our stability maps shown here are, therefore, 
accurate to within the constraint highlighted by that study.}. In total we 
ran $\sim 70,000$ simulations which required  $\sim 35,000$ hours of CPU time.

 We use the definition of the ``eccentric habitable zone'' 
(EHZ; \cite{Barnes2008}),
 which is  the HZ from Selsis et al
(2007), with $50 \%$ cloud cover,
but assumes 
the orbit-averaged flux 
 determines surface temperature (Williams \& Pollard 2002).
In other words, the EHZ is the range of orbits for which a planet receives as much flux over an orbit as
 a planet on a circular orbit in the  HZ of \cite{Selsis2007}.

\section{Results: Dynamical Stability in and around Habitable Zones}
\label{sec3}

\subsection{ Jupiter mass planet with hypothetical Earth mass planet}
\label{sec3.1}
In Figs. 1 \& 2, we show representative results of our numerical
 simulations from the
 Jupiter mass planet with hypothetical Earth mass planet 
case discussed in Section \ref{sec2}. 
 In all panels of Figs. 1 \& 2,
 the blue squares and red triangles
 represent Lagrange stable and unstable simulations respectively,
 the black circle represents the ``known''
 planet and the shaded green region represents the EHZ. For each case,
we also plot  $\bbc$ contours calculated from Eq. (\ref{bbc}).  
In any given panel, as $a$ increases, the  curves change
from all unstable (all red triangles) to all stable
  (all blue squares), with a transition region in between.

We designate a particular 
$\bbc$ contour as $\tau_\mathrm{s}$, beyond which (larger values)
 a hypothetical terrestrial mass
planet is stable for all values of $a$ and $e$, 
 for at least $10^{7}$ years.
   We tested $\tau_\mathrm{s}$ is the first $\bbc$ 
(close to the known massive planet)  that is completely stable 
(only blue squares).
 For $\bbc$ curves below $\tau_\mathrm{s}$,
 all or some locations along those curves may be unstable; 
 hence, $\tau_\mathrm{s}$ is a conservative 
representation of the Lagrange stability boundary. Similarly, we designate 
$\tau_\mathrm{u}$ as the largest value of $\bbc$ for which
all configurations are unstable. Therefore, the range
$\tau_{u} < \bbc < \tau_{s}$ is a transition region,
 where the hypothetical 
planet's orbit changes from unstable ($\tau_\mathrm{u}$)
 to stable ($\tau_\mathrm{s}$). Typically this transition
occurs over $10^{-3} \bbc$.
Although Figs. 1 \& 2 only show curves in this transition region,
 we performed many more integrations
at larger and smaller values of $\bbc$, but exclude them from the plot to
improve the readability. For all  cases, {\it all our  simulations
 with $\bbc > \tau_\mathrm{s}$ are stable, and all 
with $\bbc < \tau_\mathrm{u}$ are unstable.}

 These figures show that the Lagrange stability boundary significantly diverges
from Hill stability boundary, as the eccentricity of the known Jupiter-mass
 planet increases.
 Moreover, $\tau_\mathrm{s}$ is more or less independent
 (within $0.1 \%$)
 of whether the Jupiter mass planet lies at
$0.25$ AU or at $4$ AU.  If an extra-solar planetary system 
is known to have a Jupiter mass planet, then one can calculate $\bbc$  
over a range of $a$ \& $e$, and those regions with $\bbc > \tau_\mathrm{s}$ are
stable. We show explicit examples of this methodology in Section 4.

We also consider host star masses of $0.3 M_\odot$ and performed additional
simulations. We do not show our results here, but  they indicate 
the  mass of the star 
does not effect stability boundaries.

\subsection{ Lagrange stability boundary as a function of planetary mass \& eccentricity.}
\label{analtau}
\begin{figure}[!hbp|t]
\subfigure[] {
\includegraphics[width=.50\textwidth]{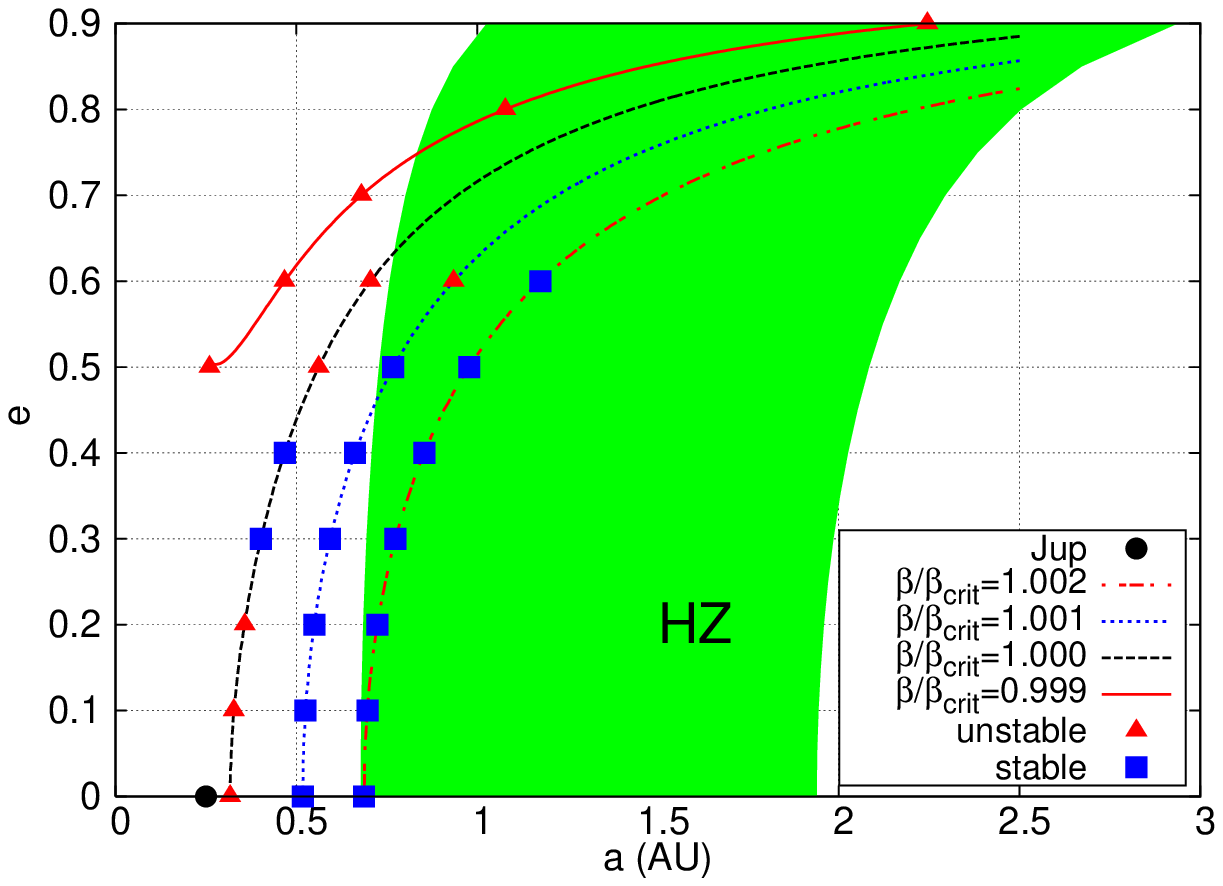}
}
\subfigure[] {
\includegraphics[width=.50\textwidth]{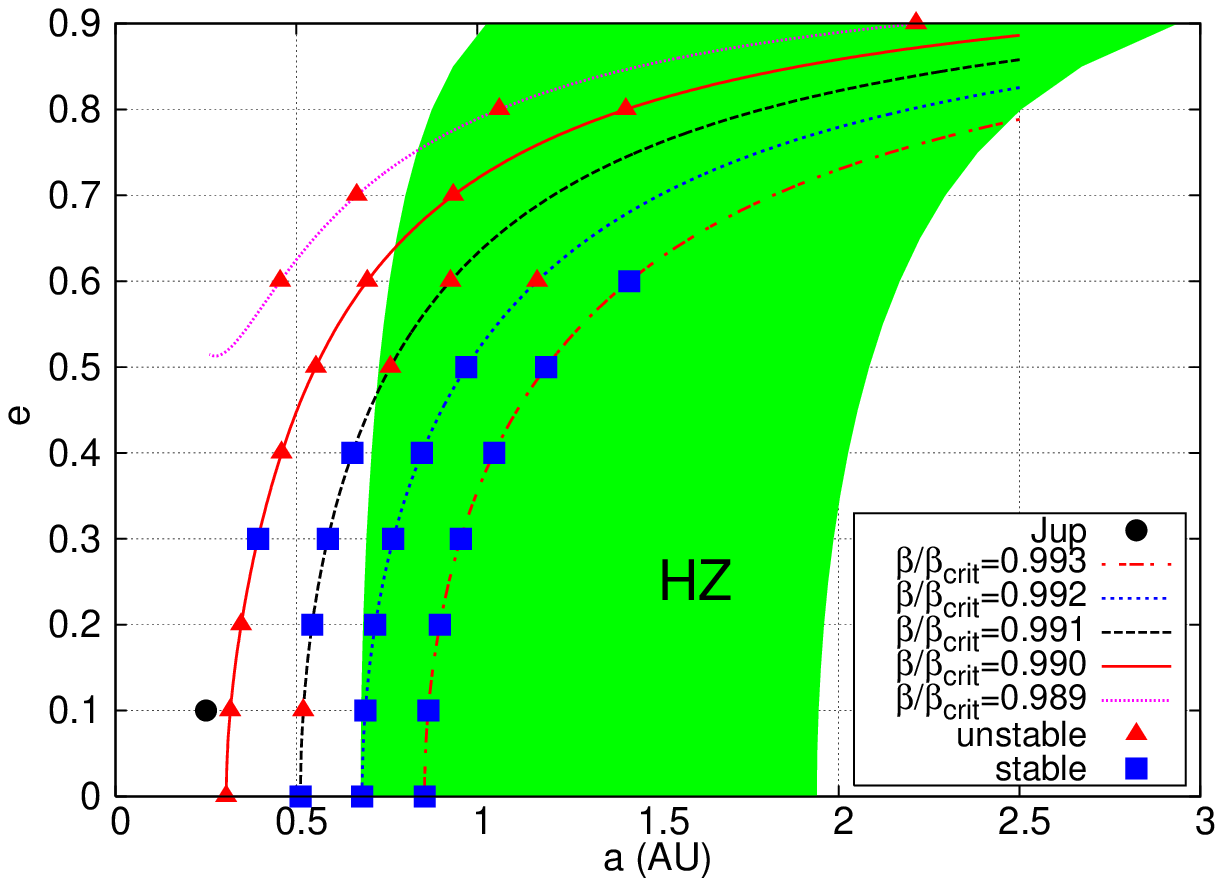}
}
\subfigure[] {
\includegraphics[width=.50\textwidth]{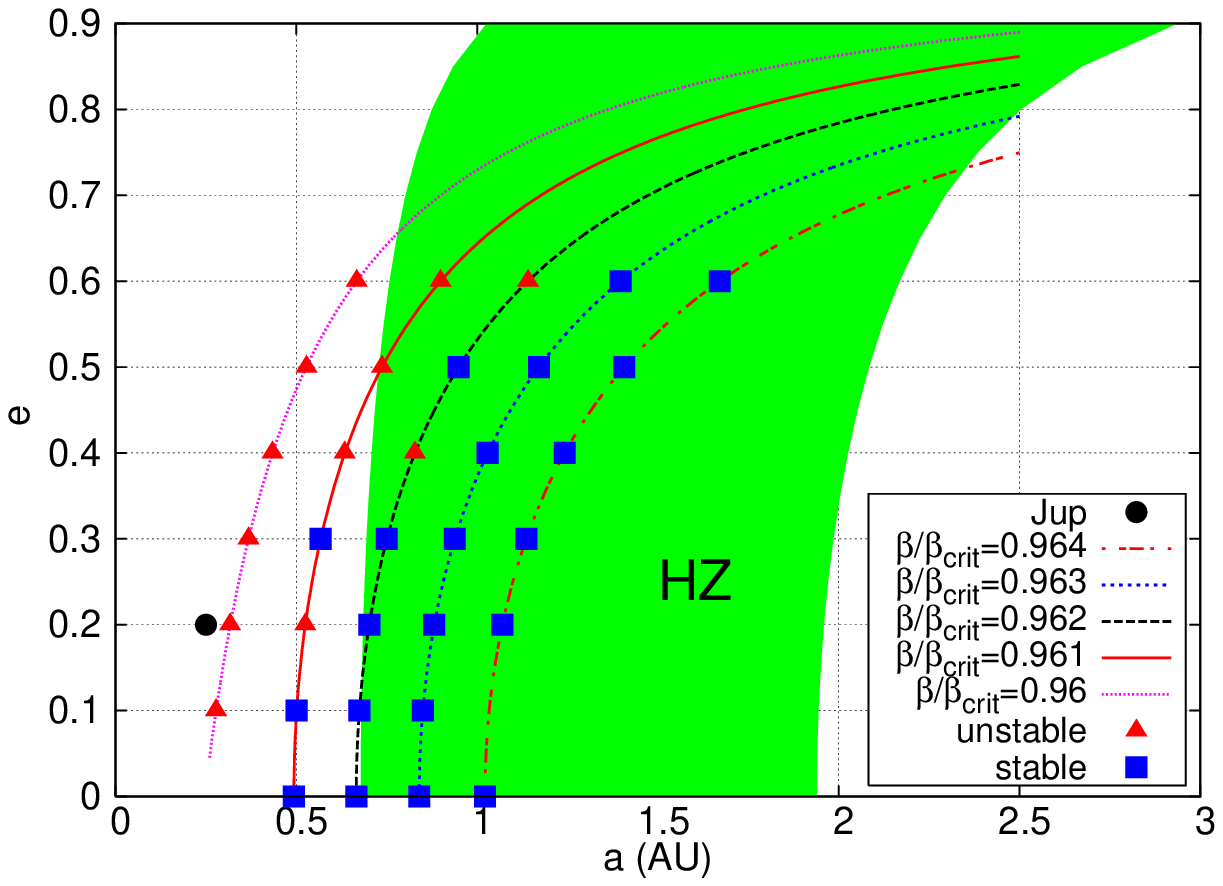}
}
\subfigure[] {
\includegraphics[width=.50\textwidth]{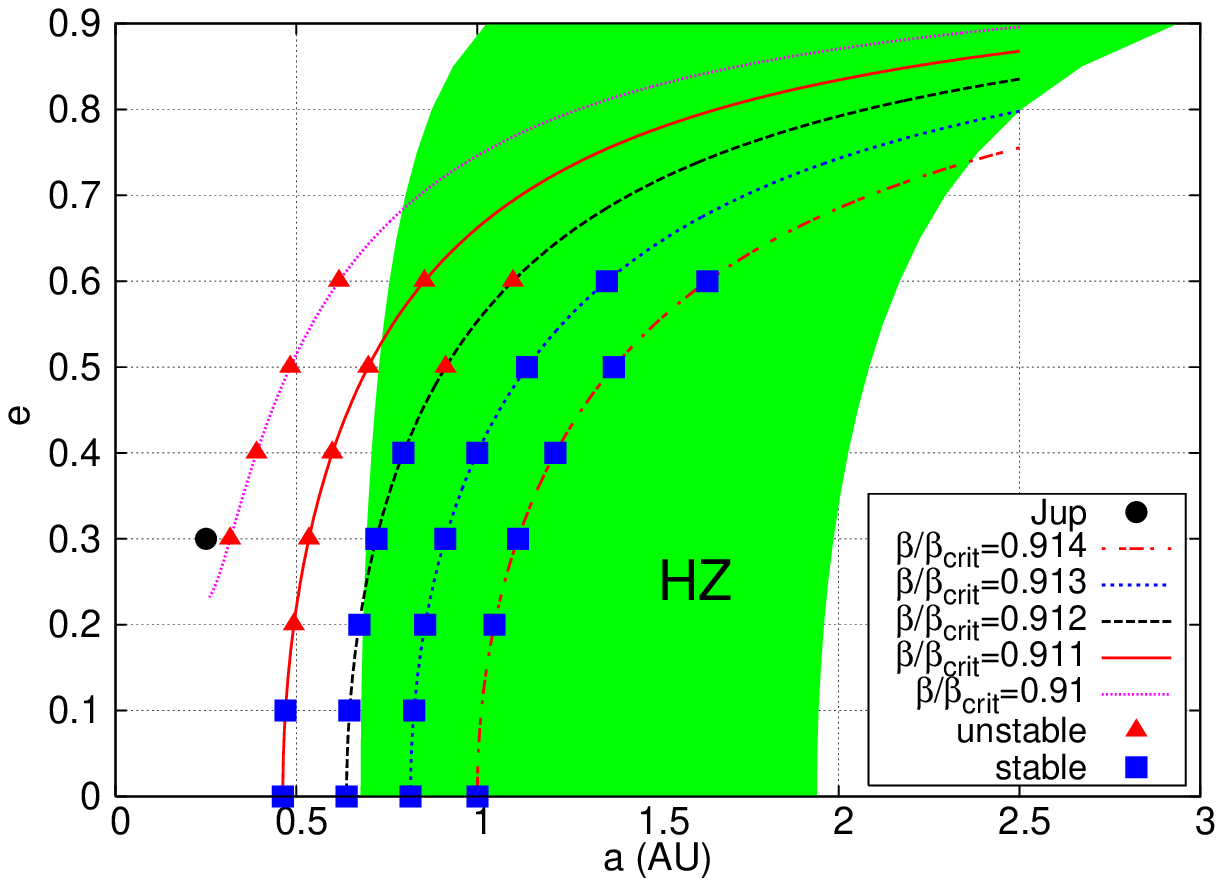}
}
\subfigure[] {
\includegraphics[width=.50\textwidth]{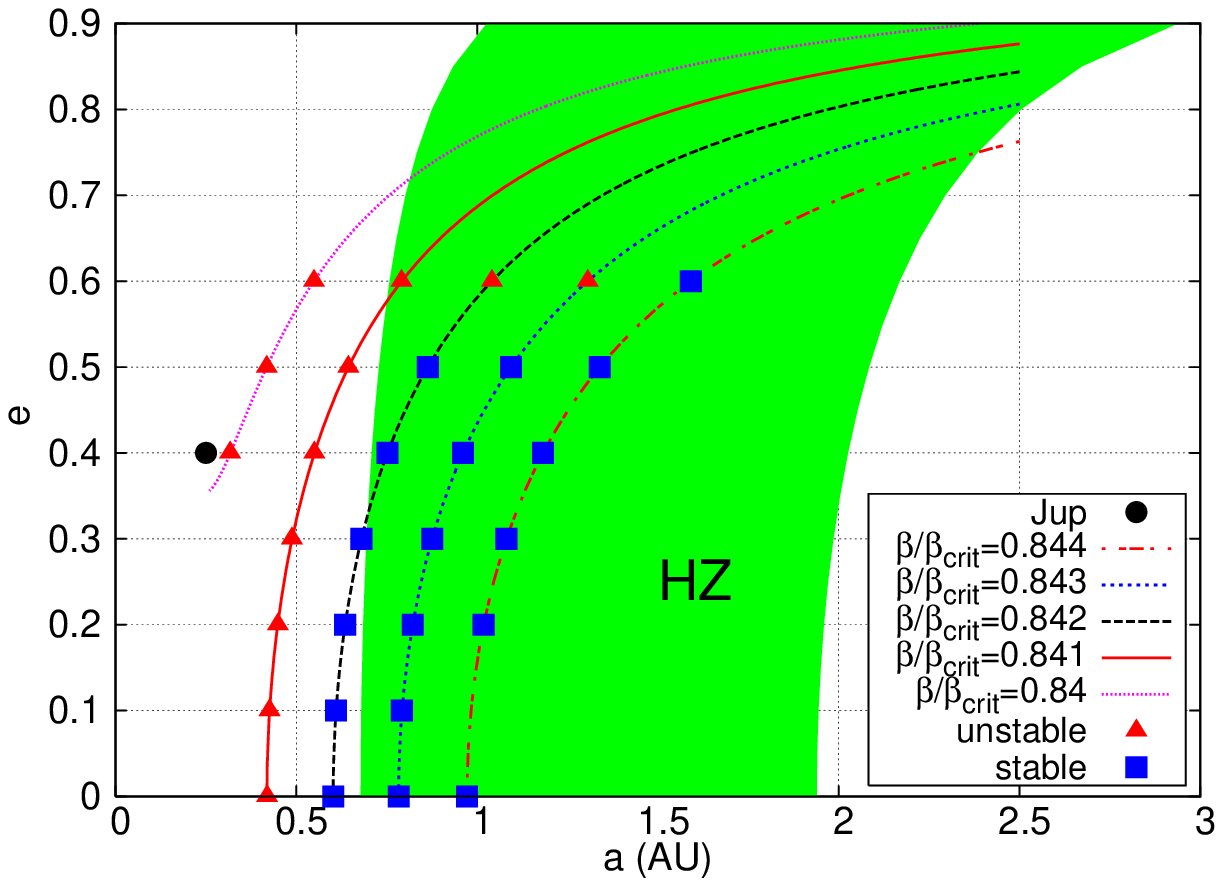}
}
\subfigure[] {
\includegraphics[width=.50\textwidth]{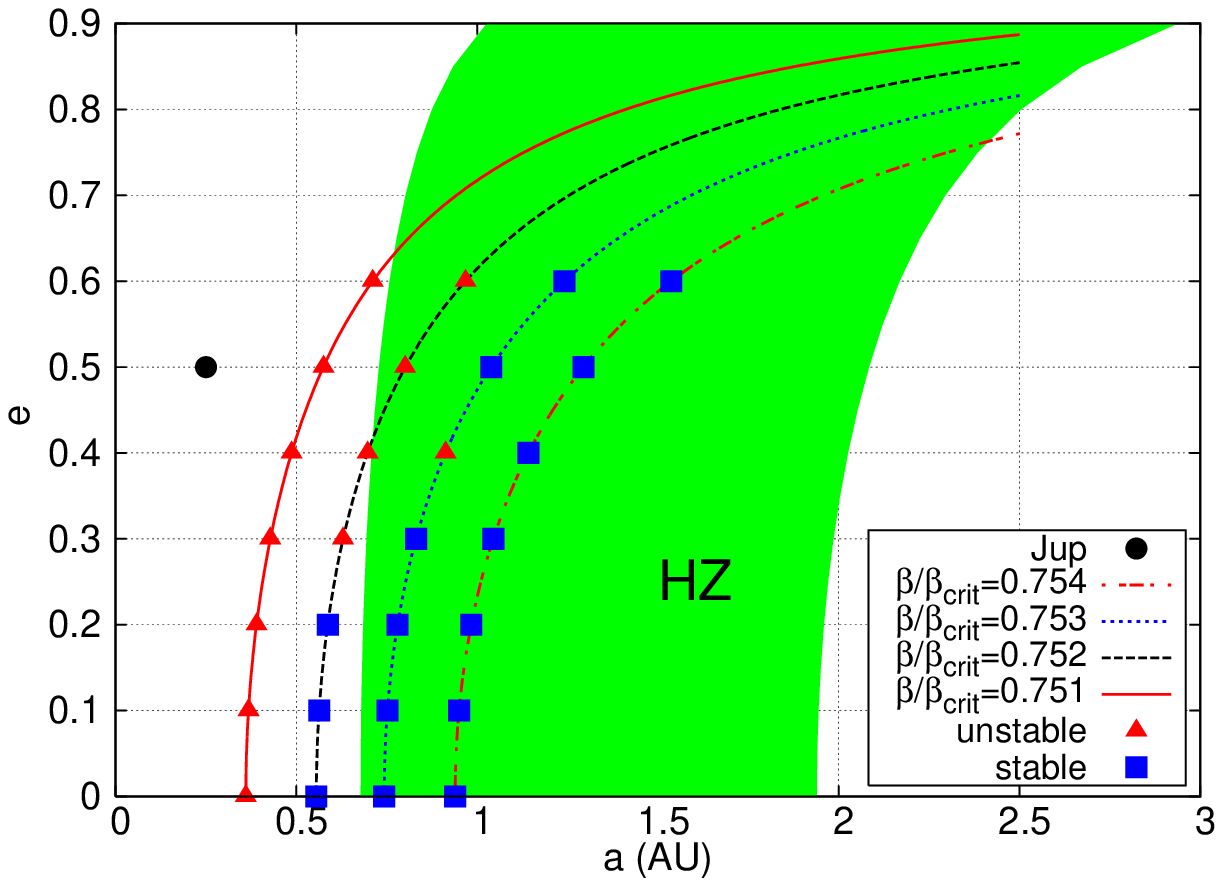}
}
\caption{ A comparison of Hill and Lagrange stability. Colored curves (shown also
in different linestyles) are 
contours of $\bbc$ (the Hill boundary lies at $\bbc = 1$). Points on the curve
designate $N-$ body simulations: red points were unstable, blue stable for
an Earth mass planet. The
green shaded region represents the HZ, and the black point is the ``known''
Jupiter mass  planet.
 The left-most curves with no stable configurations correspond to
$\tau_{u}$, the right-most that are fully stable represent $\tau_{s}$. 
(Note that for these cases we consider eccentricities $> 0.9$ in order to
identify $\tau_{u}$). }
\label{JEfig_in}
\end{figure}

\begin{figure}[!hbp|t]
\subfigure[] {
\includegraphics[width=.50\textwidth]{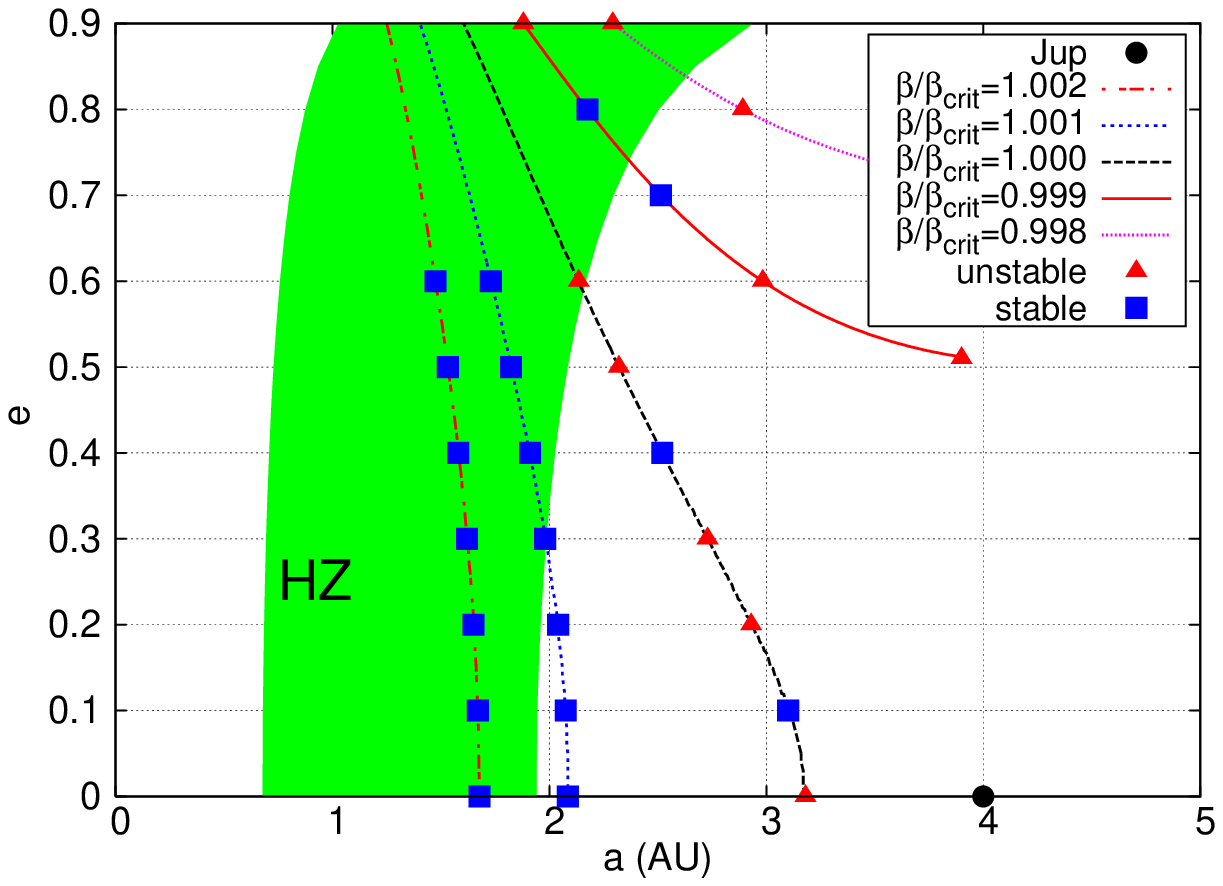}
}
\subfigure[] {
\includegraphics[width=.50\textwidth]{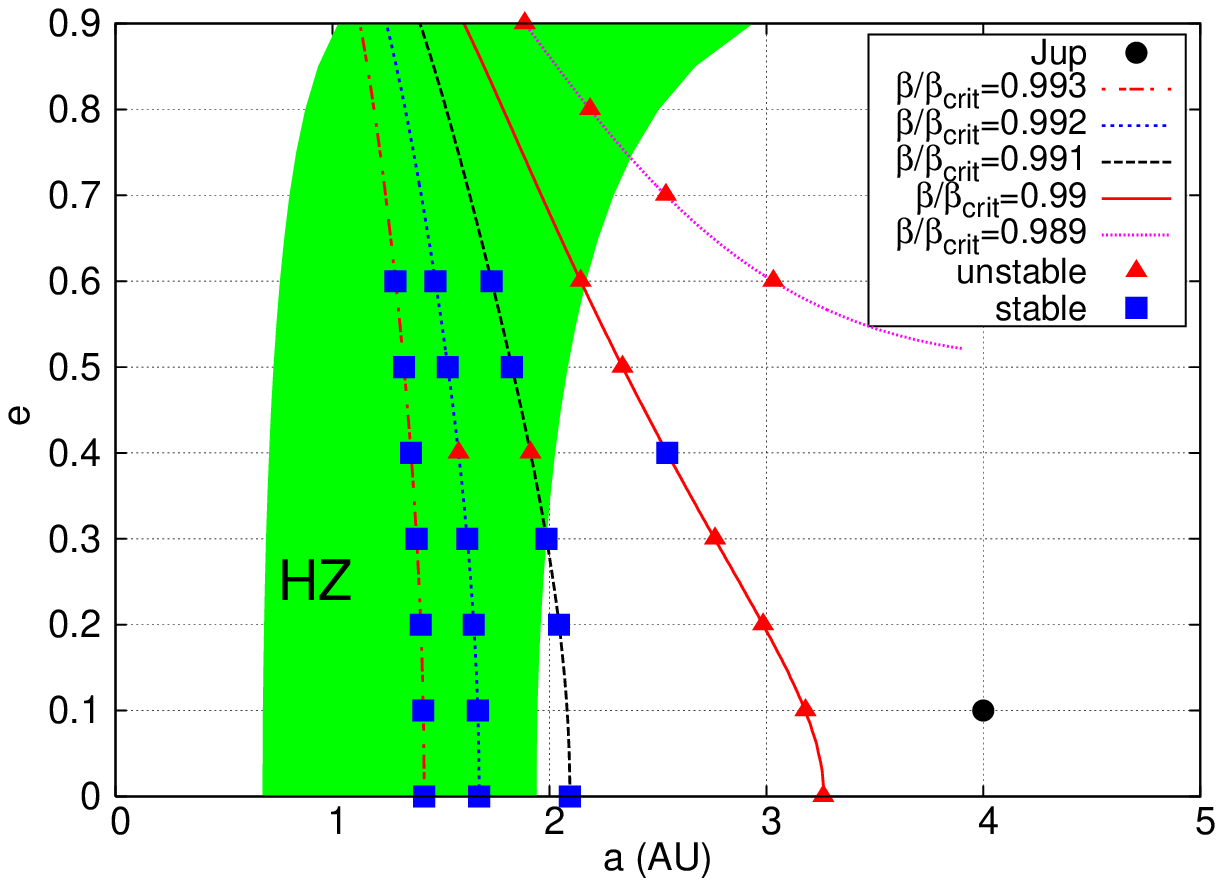}
}
\vspace{0.3in}
\subfigure[] {
\includegraphics[width=.50\textwidth]{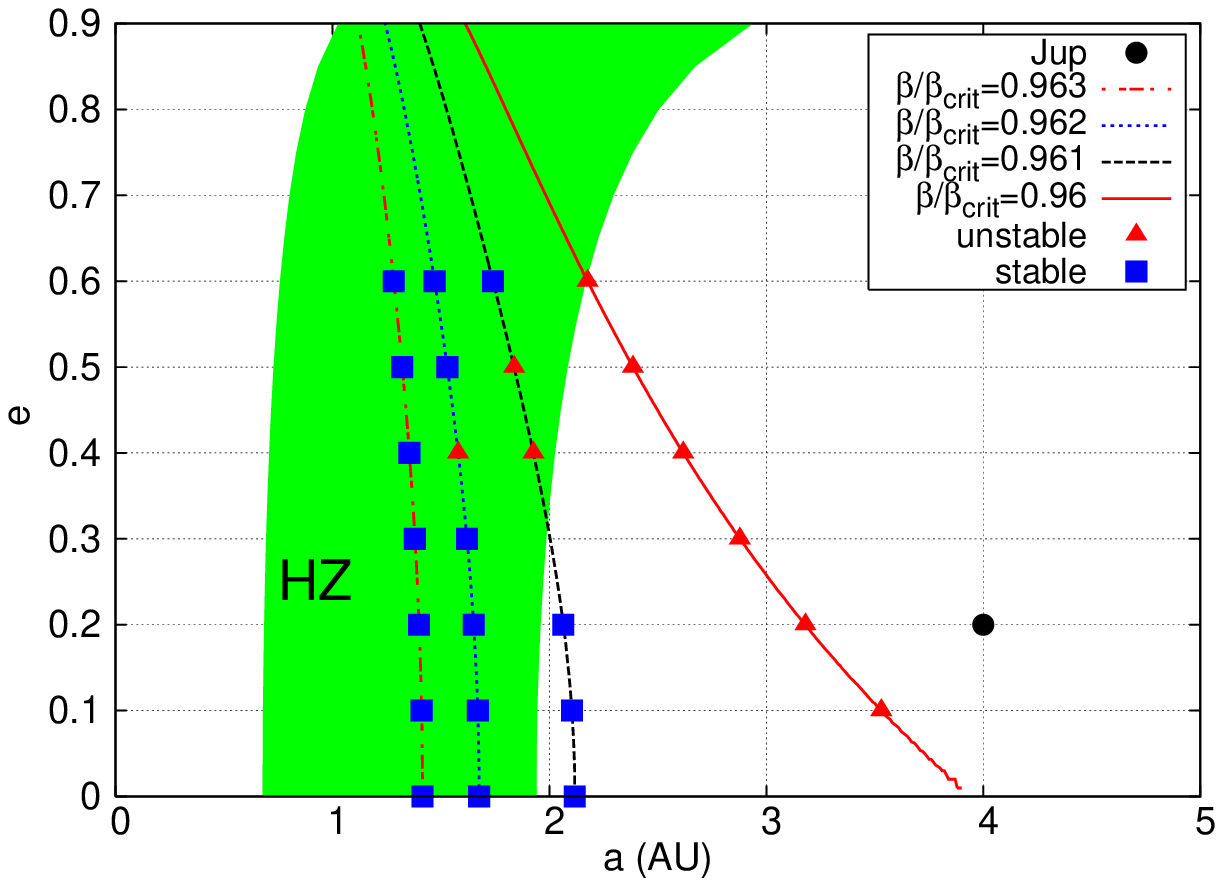}
}
\subfigure[] {
\includegraphics[width=.50\textwidth]{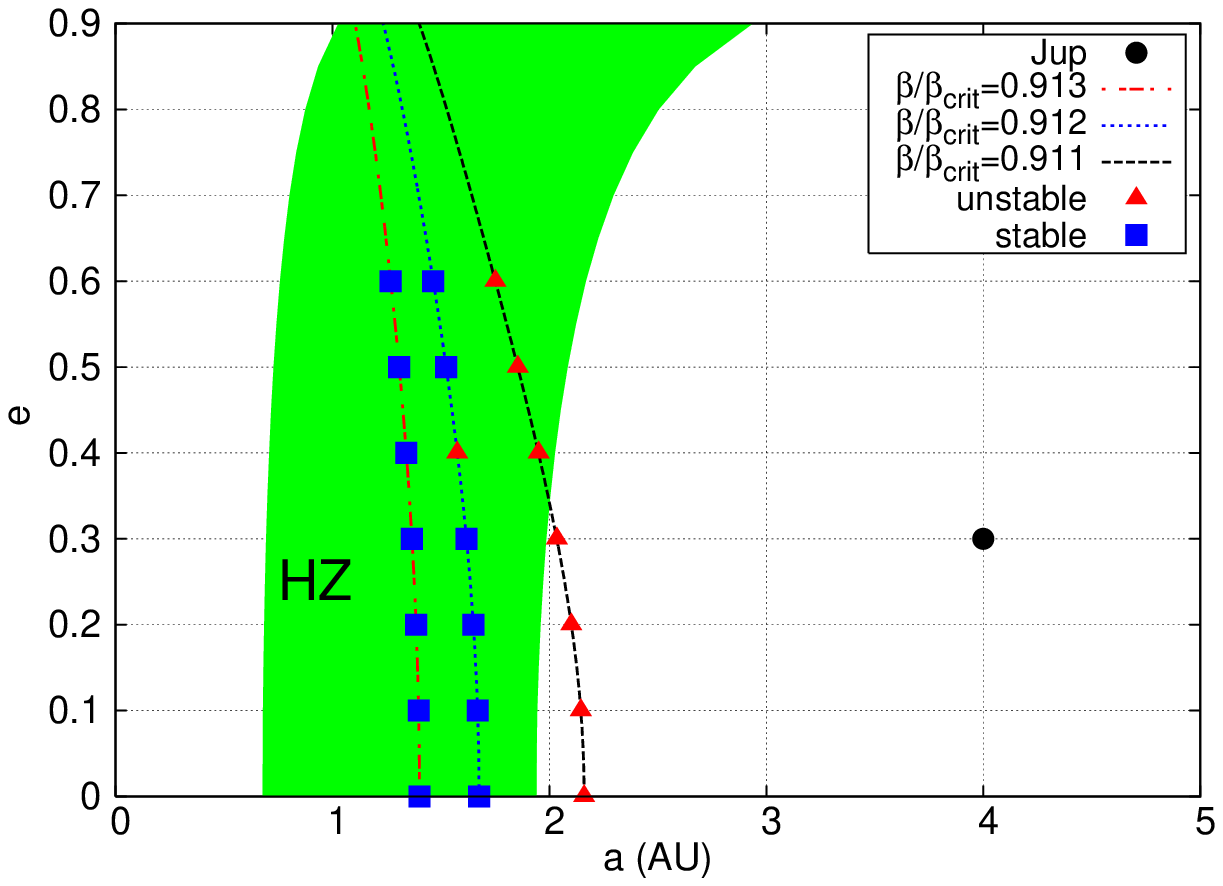}
}
\subfigure[] {
\includegraphics[width=.50\textwidth]{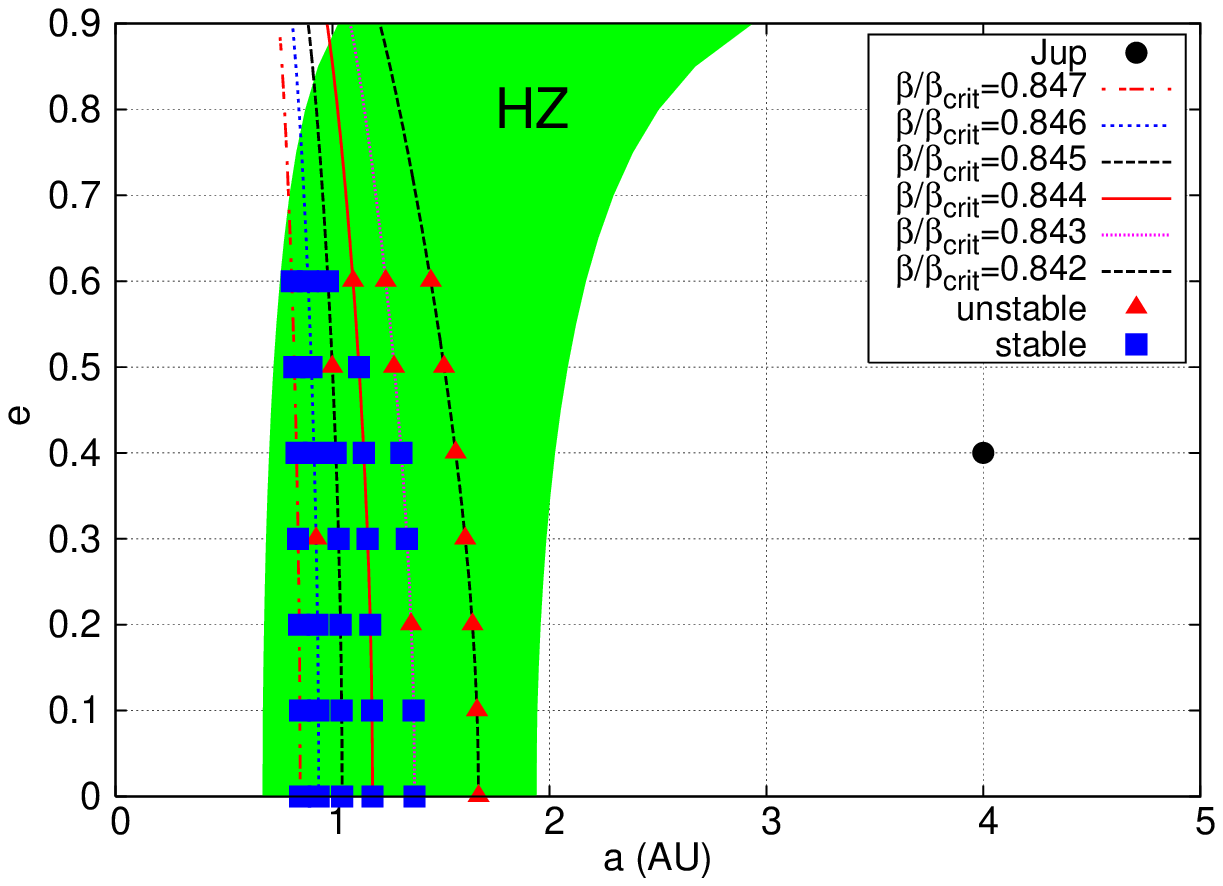}
}
\subfigure[] {
\includegraphics[width=.50\textwidth]{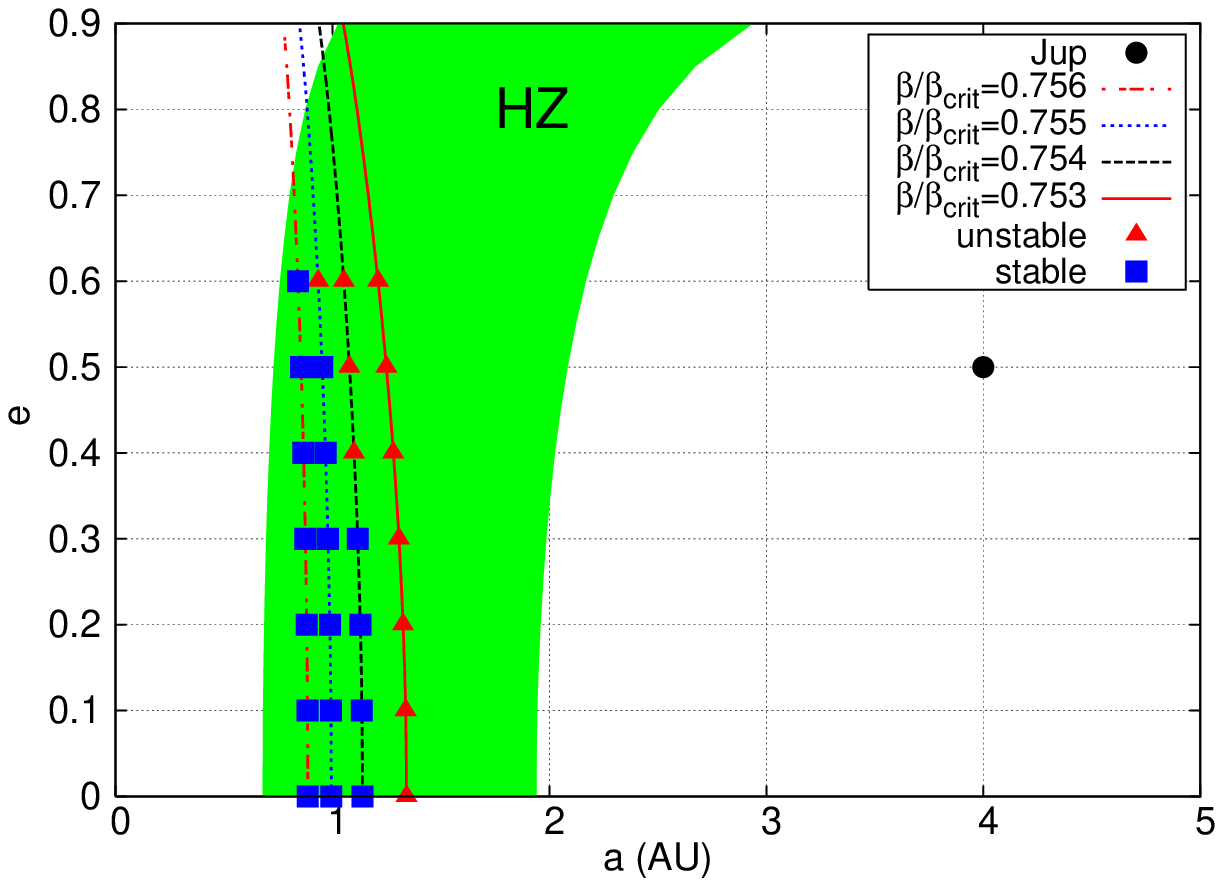}
}

\caption{ Same as  Fig. \ref{JEfig_in}, only now the Jupiter mass
planet is  at 4 AU. 
The range of values of $\tau_\mathrm{s}$ is approximately the 
same as in Fig. \ref{JEfig_in}:
from $1.002 - 0.756$.}
\label{JEfig_out}
\end{figure}

In this section we consider the broader range of ``known'' planetary masses 
discussed in Section 2 and listed in Table \ref{table1}.
Figures \ref{JEfig_in} \& \ref{JEfig_out} show that as the 
eccentricity of the ``known'' planet increases, $\tau_\mathrm{s}$ and 
$\tau_\mathrm{u}$ appear to change monotonically. This trend is apparent on 
all our simulations, and suggests $\tau_\mathrm{s}$ and 
$\tau_\mathrm{u}$ may be 
described by an analytic function of the mass and eccentricity of 
the known planet. Therefore, instead of plotting the results from these 
models 
in $a-e$  space, as shown in Figs. \ref{JEfig_in} \& \ref{JEfig_out},
we identified these analytical expressions that relate $\tau_\mathrm{s}$ 
and 
$\tau_\mathrm{u}$ to
 mass $m_{1}$ and eccentricity $e_{1}$
of the known massive planet. Although these fits were made for planets 
near the host star, these fits should apply in all cases, irrespective of 
it's distance from the star. In the following equations,
 the parameter $x = \log [m_{1}]$, where $m_{1}$ is expressed in Earth masses and
$y = e_{1}$. The stability boundaries for systems with hypothetical 
$1 \mearth$ and $10 \mearth$ mass companion are:
\begin{eqnarray}
\tau_\mathrm{j} &=& c_{1}+ \frac{c_{2}}{x} + c_{3}~y + \frac{c_{4}}{x^{2}} + c_{5}~y^{2}
+ c_{6}~\frac{y}{x} + \frac{c_{7}}{x^{3}} + c_{8}~y^{3} + c_{9}~\frac{y^{2}}{x} + c_{10}~\frac{y}{x^{2}}
\label{tauearth}
\end{eqnarray}
where $j=s,u$ indicate stable or unstable and the coefficients for each 
case are given Table 2.

\begin{figure}[!hbp|t]
\subfigure[] {
\includegraphics[angle=270,width=.56\textwidth,totalheight=0.4\textheight]{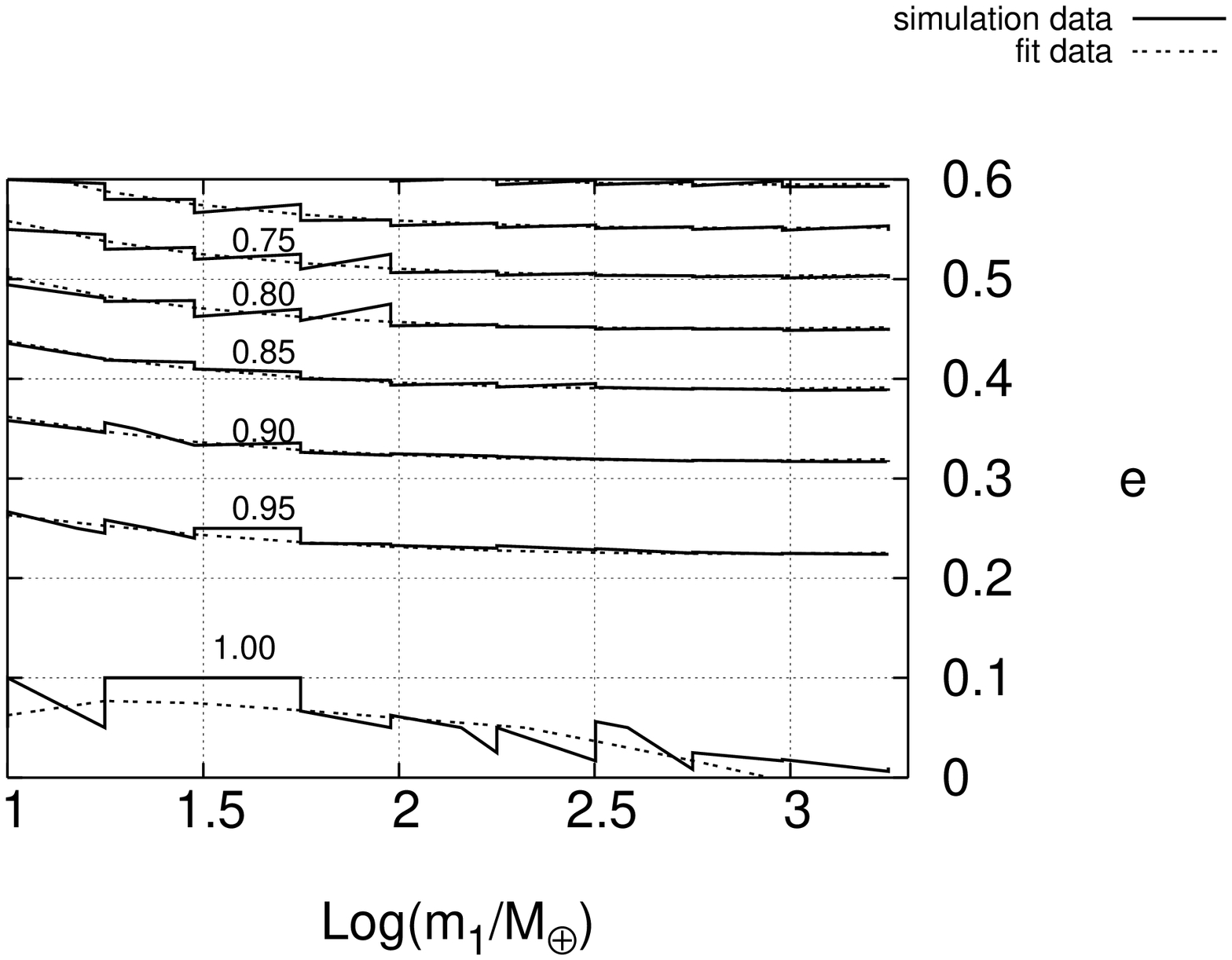}
}
\subfigure[] {
\includegraphics[angle=270,width=.56\textwidth,totalheight=0.4\textheight]{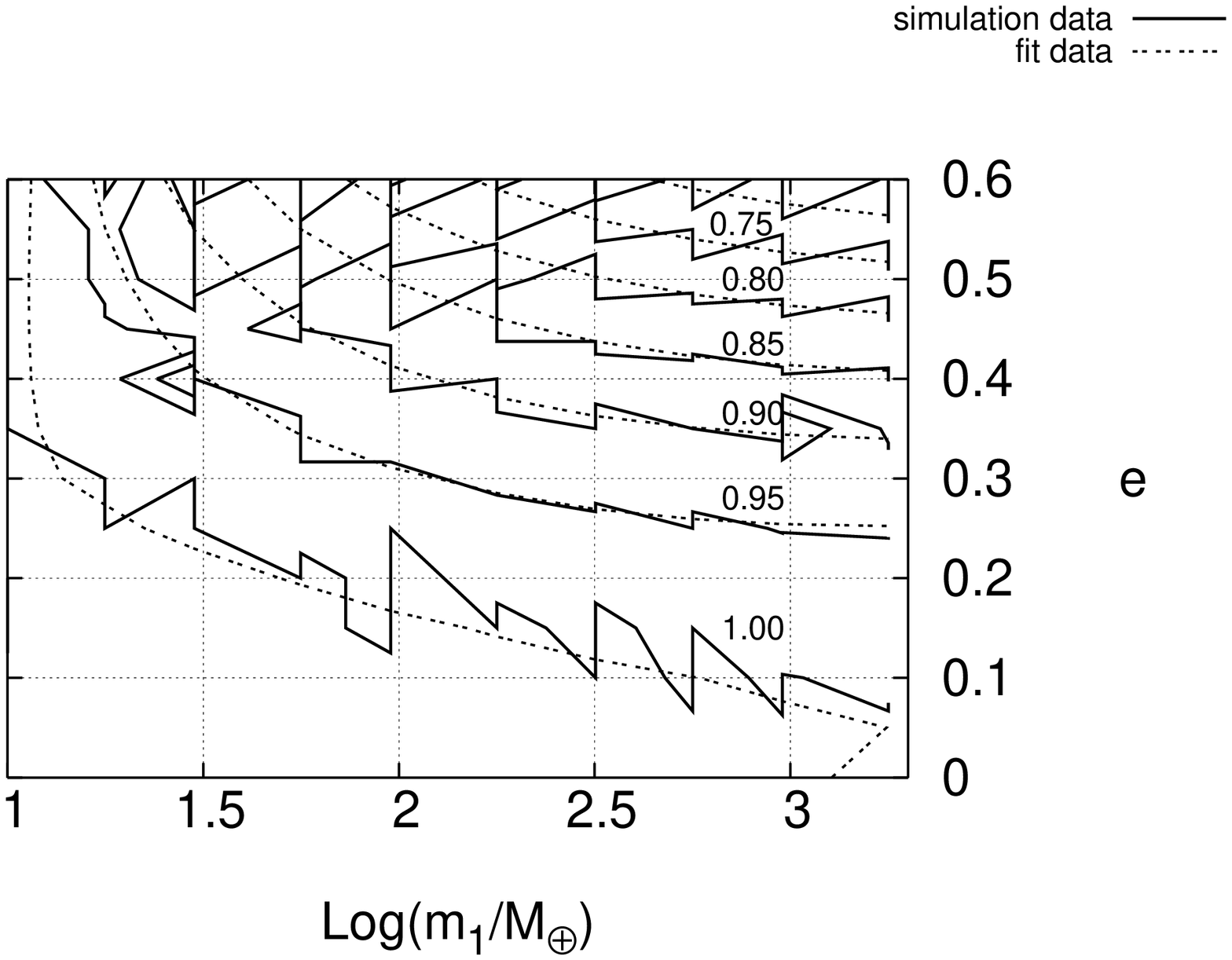}
}
\vspace{0.15in}
\subfigure[] {
\includegraphics[angle=360,width=.56\textwidth,totalheight=0.4\textheight]{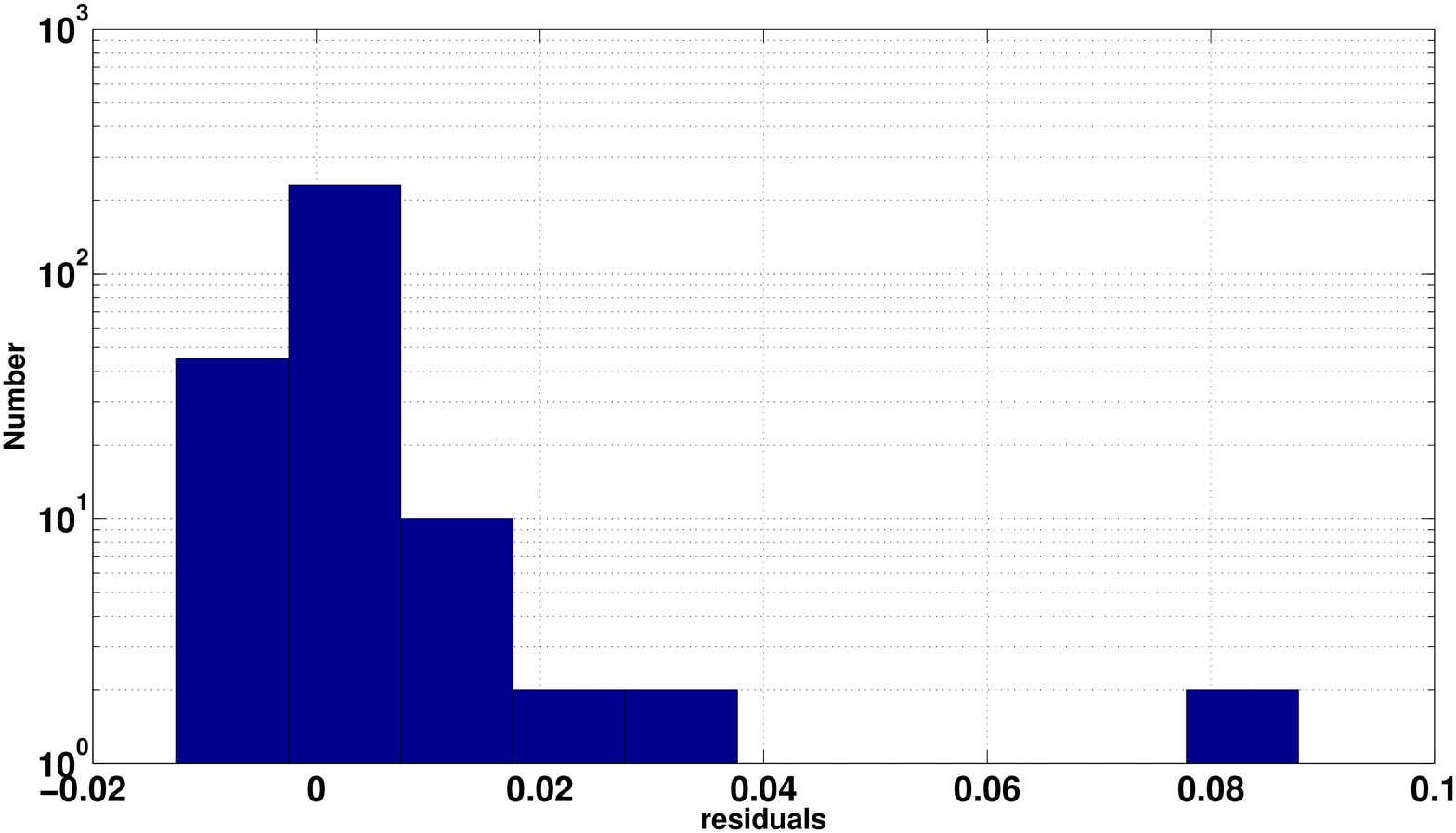}
}
\subfigure[] {
\includegraphics[angle=360,width=.56\textwidth,totalheight=0.4\textheight]{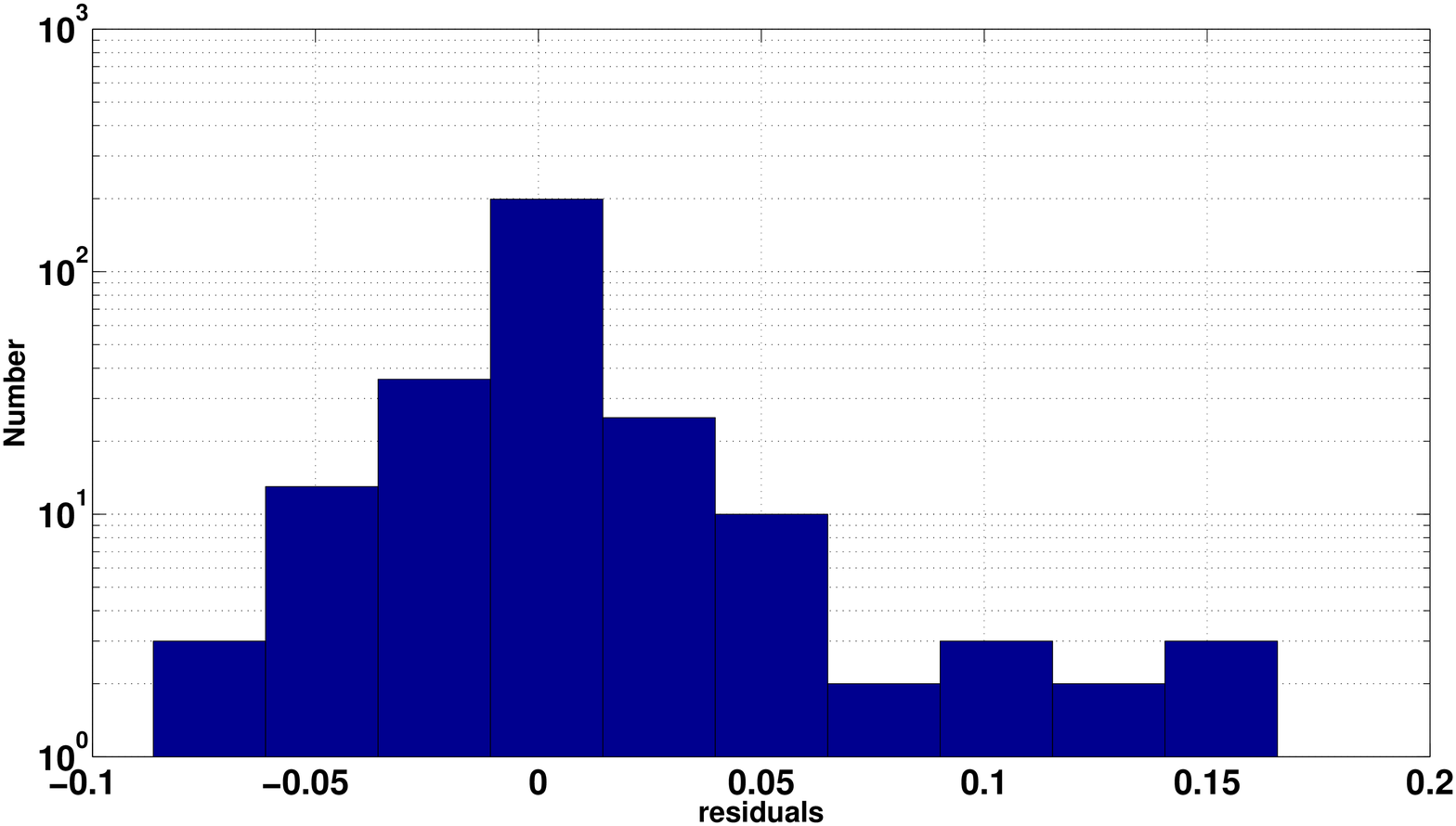}
}
\caption{In the top panels, we show contours of $\tau_\mathrm{s}$
 from numerical simulations (solid line) compared to the best fit (dashed line
) in 
$\log [m_{1}]$ -- $e$ space, for $1 \mearth $ (panel a) and $10 \mearth$
(panel b) companions. The 
expression for the best fit is given in Eq. (\ref{tauearth})
with appropriate coefficients given in Table 1. 
The bottom panels show residuals between the numerical results and the best fit,
 with a standard deviation of $0.0065$ for $1 \mearth $ companion (panel c) and
 $0.0257$ for $10 \mearth$ companion (panel d).}
\label{SHB_jup}
\end{figure}

The coefficients in the above expression were obtained by finding a best
fit curve to our model data that maximizes the  $R^{2}$ statistic, 
\begin{eqnarray}
R^{2} &=& 1 - \frac{\displaystyle\sum_{i}^n (\tau^{model}_{i} - \tau^{fit}_{i})^2}{\displaystyle\sum_{i}^n (\tau^{model}_{i} - \overline{\tau^{model}})^2}
\end{eqnarray}
where $\tau^{model}_{i}$ is the $i^{th}$model value of $\tau$ from numerical
 simulations, 
$\tau^{fit}_{i}$ is the corresponding model value from the curve fit, 
$\overline{\tau^{model}}$ is the average of all the $\tau_{model}$ values 
and $n=572$ is the number of models 
(including mass, eccentricity and locations of the massive planet).
Values close to $1$ indicate a better quality of the fit. 
In Fig. \ref{SHB_jup}, the top panels (a) $\&$ (b) show contour maps of 
$\tau_\mathrm{s}$ as a function of $\log [m_{1}]$ and $e_{1}$ between model data
(solid line) and best fit (dashed line). The  $R^{2}$ values
 for $1 \mearth$ companion 
(Fig. \ref{SHB_jup}(a)) and  $10 \mearth$ companion (Fig. \ref{SHB_jup}(b)) are
$0.99$ and $0.93$, respectively, for $\tau_\mathrm{s}$.
In both the cases, 
the model and the fit deviates  when the masses of both the planets are near 
terrestrial mass. Therefore our analysis is most robust for more 
unequal mass planets.
The residuals between the model and the predicted $\tau_{s}$ values are
also shown in Fig. \ref{SHB_jup}(c) ($1 \mearth$ companion) $\&$ 
Fig. \ref{SHB_jup}(d) ($10 \mearth$ companion). The standard deviation of these
residuals is $0.0065$ and $0.0257$ for $1 \mearth$ and $10 \mearth$,
respectively, though the $1 \mearth$ case has an outlier which does not significantly effect the fit.
The maximum deviation is $0.08$ for $1 \mearth$ and $0.15$ for $10 \mearth$ cases.
The expression given in Eq. (\ref{tauearth}) can be used to identify 
Lagrange stable regions ($\bbc > \tau_\mathrm{s}$)
for terrestrial mass planets  around  stars with one known planet with $e \le 0.6$
and may provide an important tool for the observers to locate these
planets\footnote{Note that a more thorough exploration of mass and eccentricity parameter
space may indicate regions of resonances on both sides of the stability. Hence,
 we advice caution in applying our expression in those regions.}.
 Once a Lagrange stability boundary is identified, it is
 straightforward to calculate the
range of $a$ and $e$ that is stable for a hypothetical terrestrial mass planet, using
Eq.(\ref{bbc}).
 In the next section, we illustrate the
applicability of our method for selected observed systems.

\section{Application to observed systems}
\label{sec4}

The expressions for $\tau_{s}$ given in \S\ref{analtau} can be very useful 
in calculating parts of HZs that are
 stable for all currently known single planet systems.
 In order to calculate this fraction, we used 
orbital parameters from the Exoplanet Data Explorer maintained by 
California Planet Survey consortium\footnote{\url{http://exoplanets.org/}},
and selected all $236$ single planet systems  in this database with masses 
in the range
$10 \mjup$ to $10 \mearth$ and $e \leq 0.6$.

 
Table \ref{table2} lists the properties of the example systems that 
we consider in \S\ref{rhocrb}-\S\ref{hd7924}, along with the 
orbital parameters of the known companions and stellar mass. The procedure
to determine the extent of the stable region for a hypothetical 
$1 \mearth$ and $10 \mearth$ is as follows:
(1) Identify the mass ($m_{1}$) and eccentricity ($e$) of the known planet.
(2) Determine $\tau_\mathrm{s}$ from Eq. \ref{tauearth} with coefficients from Table 2.
(3) Calculate $\bbc$ over the range of orbits ($a$ and $e$) around the
known planet using Eq. \ref{bbc}.
(4) The Lagrange stability boundary is the $\bbc = \tau_{s}$ curve.

\subsection{Rho CrB}
\label{rhocrb}
As an illustration of the internal Jupiter + Earth case, we consider the Rho CrB system.
Rho CrB is a G0V star with a mass similar to the
Sun, but with greater luminosity.
 \cite{Noyes1997} discovered  a
 Jupiter-mass  planet orbiting at
 a distance of $0.23$ AU with low eccentricity ($e=0.04$).
Since the current inner edge of the circular HZ of this star lies
 at $0.90$ AU, there is a good possibility for terrestrial planets to 
remain stable within the HZ. Indeed, \cite{Jones2001} found that stable orbits
may be prevalent in the present day {\it circular} HZ of Rho CrB for Earth mass planets.

\begin{figure}[!hbp|t]
\subfigure[]{
\includegraphics[width=0.53\textwidth]{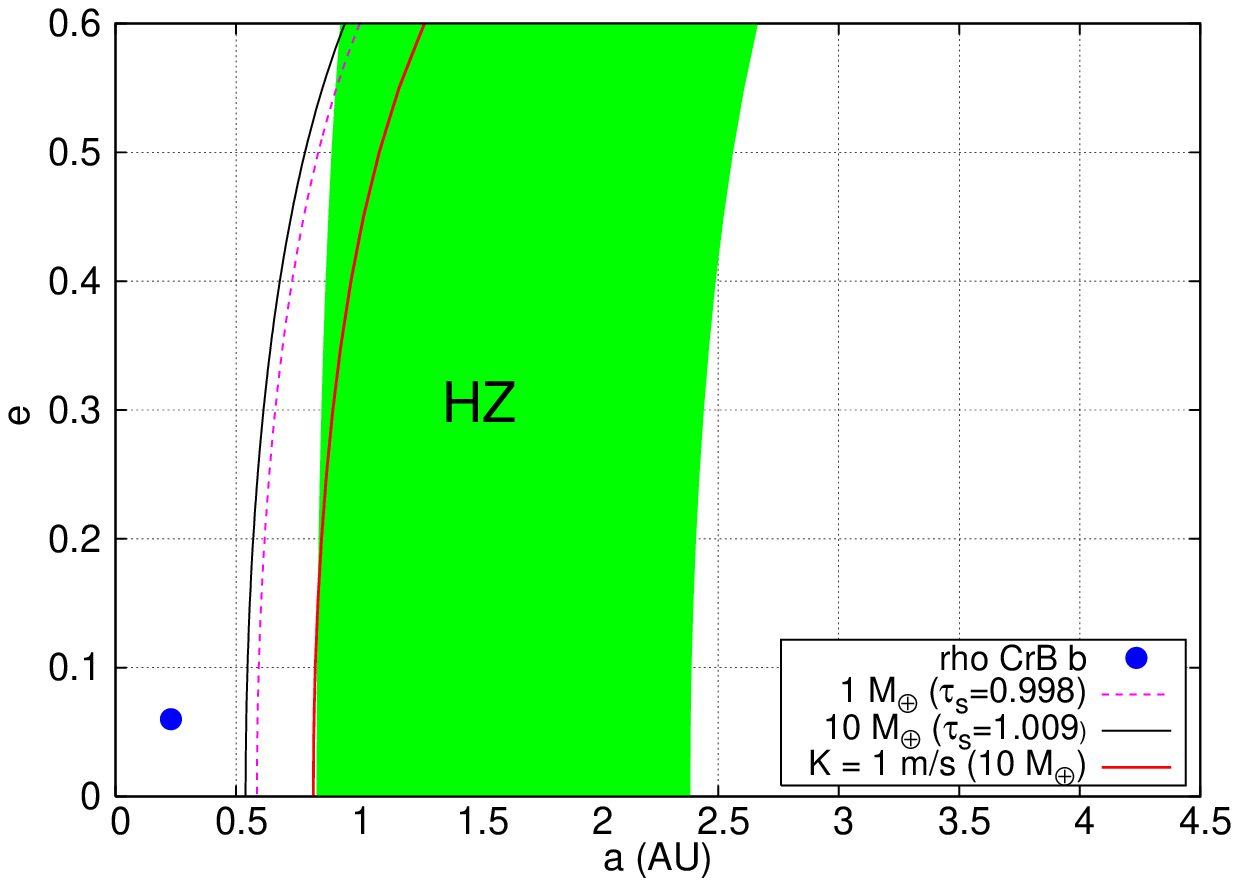}
}
\subfigure[] {
\includegraphics[width=0.53\textwidth]{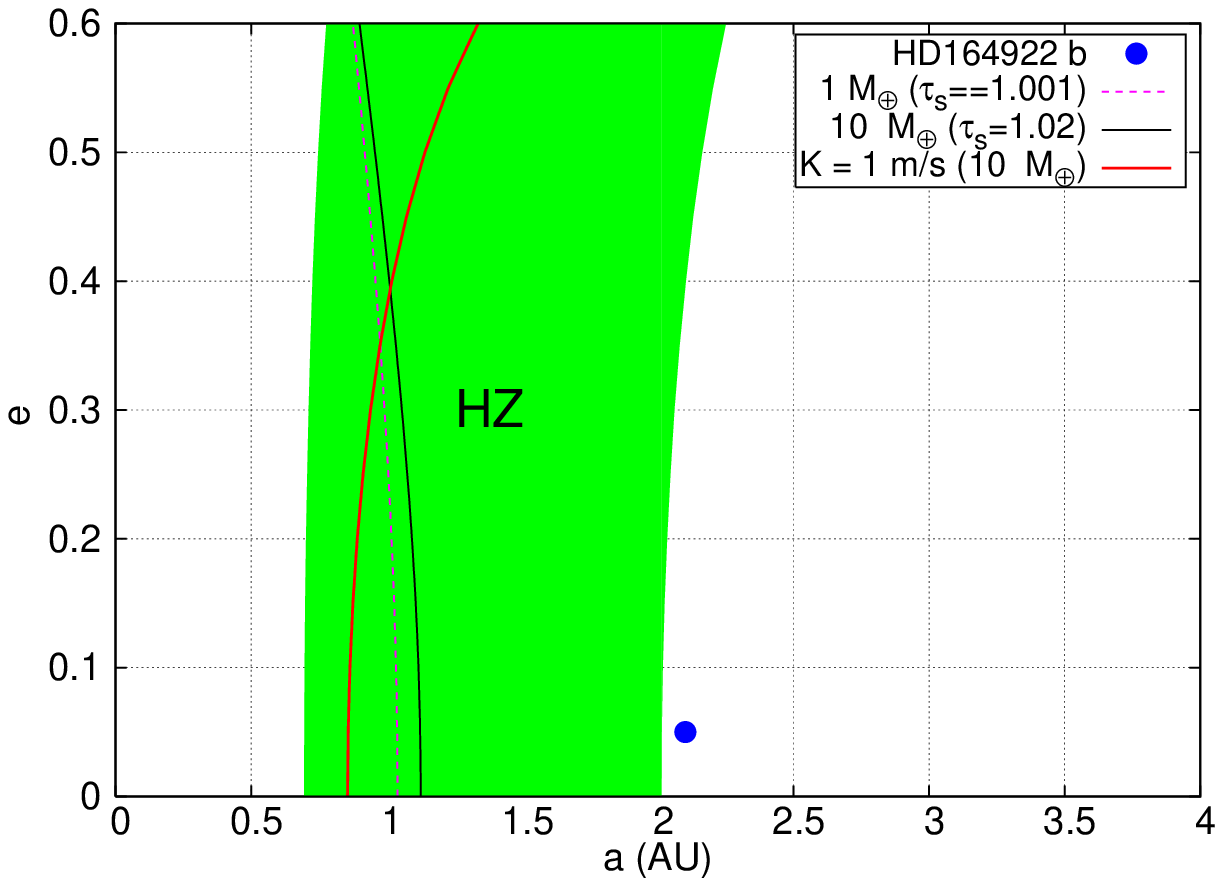}
}
\vspace{0.3in}
\subfigure[]{
\includegraphics[width=0.53\textwidth]{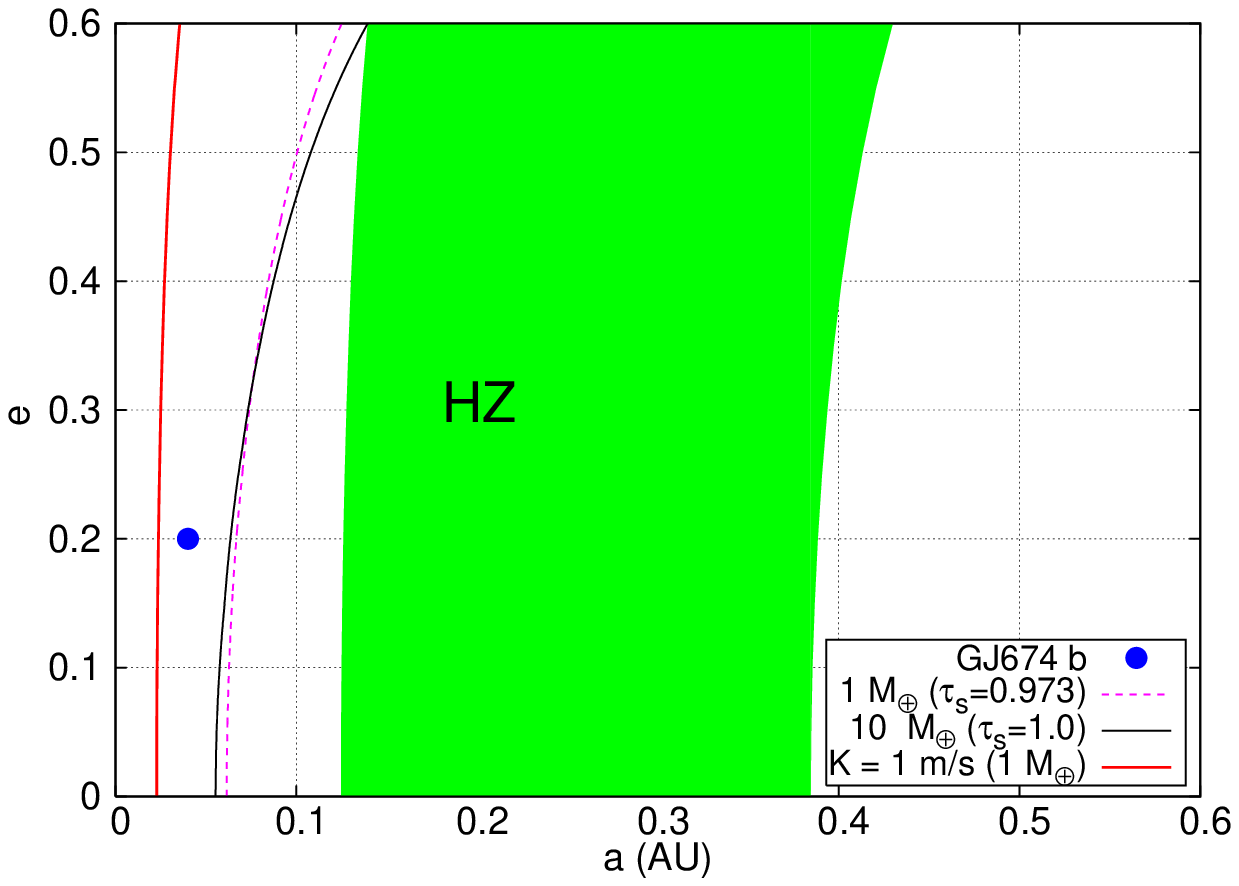}
}
\subfigure[] {
\includegraphics[width=0.53\textwidth]{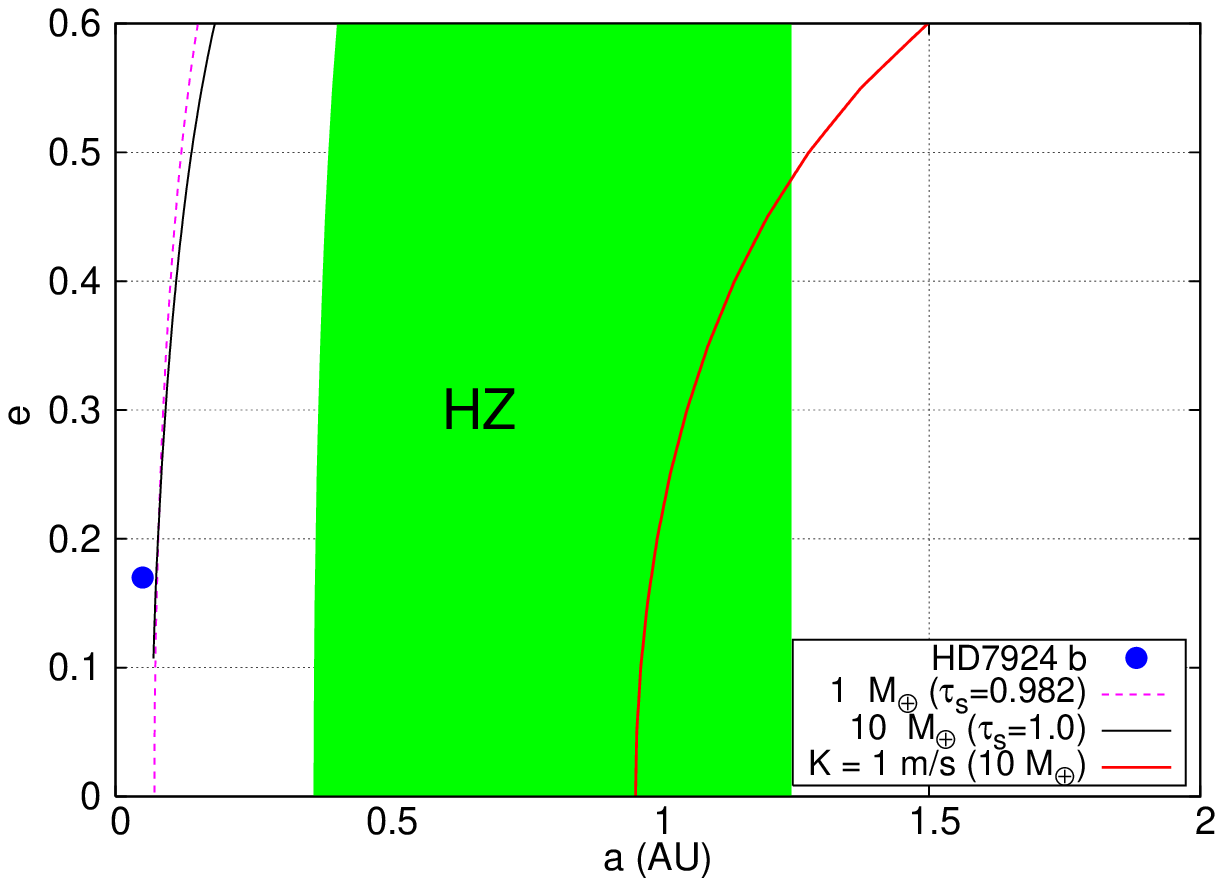}
}
\caption{Comparison of Lagrange stable regions and habitable zones for four known
systems. The magenta curves represent $\tau_{s}$ for a $1 \mearth$ 
planet, black $10 \mearth$ (c.f. Fig. 3). For panels $a$, $c$ and $d$,
stable orbits lie to the right of these curves, but lie to the left 
in panel $b$. The red solid line shows the $1$ ms$^{-1}$ RV amplitude
of a hypothetical terrestrial planet on a circular orbit. The green
 region is the HZ. }
\label{observed}
\end{figure}

 Fig. \ref{observed}a  shows the EHZ (green shaded)
 assuming a 50\% cloud cover in the $a-e$ space of Rho CrB.
The Jupiter mass planet is the blue filled circle.
Corresponding $\tau_\mathrm{s}$ values calculated from Eq.(\ref{tauearth})
 for $1 \mearth$ companion
 ($0.998$, dashed magenta line)
 and $10 \mearth$ companion ($1.009$, black solid line) are also shown.
These two contours represent the stable boundary beyond which an Earth-mass
 or super-Earth
 will remain stable  for all values of $a$ and $e$
 (cf. Figs. \ref{JEfig_in}a and \ref{JEfig_in}b).
The fraction of the HZ that is stable for $1 \mearth$ is $72.2 \%$ and  
for
$10 \mearth$ is $77.0 \%$.
Therefore the Lagrange stable stable region is larger for a larger terrestrial
 planet. We  conclude that the HZ of rho CrB can support 
 terrestrial-mass planets, except for very high eccentricity
($e > 0.6$).

 These results are in  agreement with
 the conclusion of \cite{Jones2001} and \cite{MT2003}, who found
that a planet with a mass equivalent to the Earth-moon system,
 when launched with 
$e=0$ within the HZ of Rho CrB, can remain stable for $\sim 10^{8}$ years. They also
varied the mass of Rho CrB b up to $8.8 \mjup$ and still found that the HZ is 
stable. Our models also considered systems with $3 \mjup$, $5 \mjup$ 
and $10 \mjup$ and our results show that 
even for these high masses, if the initial eccentricity of the Earth-mass planet
is less than $0.3$, then it is stable.

To show the detectability of a $10 \mearth$ planet, we have also drawn a radial velocity (RV) 
contour of 1 ms$^{-1}$ (red curve),
which indicates that  a $10 \mearth$ planet in the 
HZ  is  detectable. A similar contour
for an Earth mass planet is not shown because the precision required  is 
extremely high.

\subsection{ HD 164922}
\label{hd164922}

\cite{Butler2006} discovered a Saturn-mass planet ($0.36  \mjup$) orbiting 
HD  164922  with a period of 1150 days ($a = 2.11$ AU) and an eccentricity of
$0.1$. Although it has a low eccentricity, the uncertainty (0.14)
 is larger than the value itself. Therefore, the appropriate 
 Saturn mass cases could legitimately use any $e$ in the range $0.0 < e < 
0.25$ cases, but we use $e=0.1$.

Figure \ref{observed}b, shows the stable regions in
 the EHZ (green shaded)
 of HD 164922, for hypothetical Earth (magenta) and super-Earth 
(black) planets. The Saturn-mass planet (blue filled circle) is also
shown at 2.11 AU. About $28 \%$ of the
HZ in HD 164922 is stable for a $10 \mearth$ planet 
(for eccentricities $ \lsim 0.6$), whereas for Earth
mass planets only $10 \%$ of the HZ  is stable.
 We again show the detection limit for a $10 \mearth$
case.

\subsection{ GJ 674}
\label{gj674}

GJ 674 is an M-dwarf star with a mass of $0.35 M_\odot$ and an effective temperature of
$3600 $ K. \cite{bonfils2007} found
a  $12 \mearth$ with an orbital period and eccentricity 
of $4.69$ days 
($a = 0.039$ AU) and $0.20$, respectively.
  Fig. \ref{observed}c
shows the EHZ of GJ 674
in $a - e$ space. Also shown are the known planet GJ 674 b (filled blue
circle), EHZ (green shaded), and detection limit for an Earth-mass
planet (red curve). 
The values of $\tau_\mathrm{s}$ for $1 \mearth$ and $10 \mearth$ planets,
from Eq.\ref{tauearth}, are $0.973$ (magenta) and
$1.0$ (black), respectively.
Notice that the fraction of the HZ that is stable for $1 \mearth$ is
 slightly greater ($79.1 \%$) than 
$10 \mearth$ planet ($78.8 \%$), which differs from previous systems we considered here. 
A similar behavior can be seen in another system (HD 7924) that is discussed in the next section.
It seems that when the planet mass ratio is approaching $1$, the HZ of a $10 \mearth$ mass
planet offers less stability at high eccentricities ($> 0.6$) than a $1 \mearth$ planet. But
as noted in \S\ref{analtau}, this analysis should be weighted with the fact
 that our fitting procedure is not as accurate for a
$10 \mearth$ planet than a $1 \mearth$ planet.

\subsection{ HD 7924}
\label{hd7924}

Orbiting a K0 dwarf star at $0.057$ AU, 
the super-Earth HD 7924 b was discovered by NASA-UC Eta-Earth survey by
the California Planet Search (CPS) group \citep{Howard2009},
 in an effort to find planets in
the mass range of $3-30 M_\oplus$. It is estimated to have an
$M \sin i = 9.26 M_\oplus$ with an eccentricity of $0.17$. Fig. \ref{observed}d
shows $\tau_\mathrm{s}$ values for  hypothetical 
 $10 M_\oplus$ (magenta) and
 $1 M_\oplus$ (black  line) planets are $1.00$ and $0.98$,
 respectively.  Unlike GJ 674, where only  part of
the HZ is stable, around $94 \%$ of  HD 7924's HZ is stable for these potential planets.
Furthermore, we have also plotted an RV contour of $1$ ms$^{-1}$ arising
from the $10 M_\oplus$ planet (red curve). This indicates that  this planet
may lie above the current detection threshold, and may even be in the HZ.

 \cite{Howard2009} do find some additional 
best-fit period solutions with very high eccentricities ($ e > 0.45$), but
combined with a false alarm probability (FAP) of $> 20\%$, they conclude that
these additional signals are probably  not viable planet 
candidates. Further monitoring  may
confirm or forbid the existence of additional planets in this system.

\subsection{Fraction of stable HZ}
\label{subfraction}
\begin{figure}[!hbp]
\subfigure[]{
\includegraphics[width=0.92\textwidth]{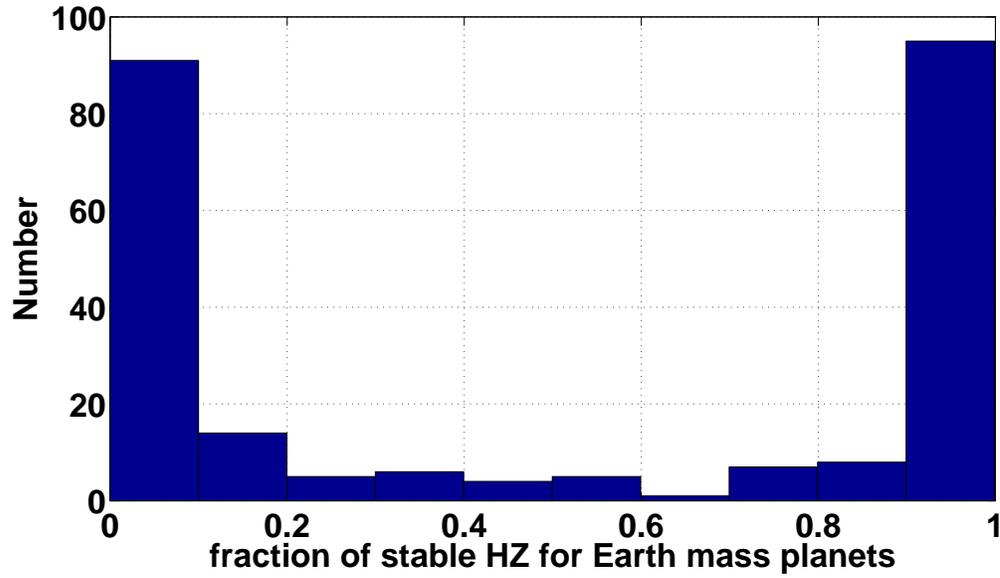}
}
\subfigure[]{
\includegraphics[width=0.92\textwidth]{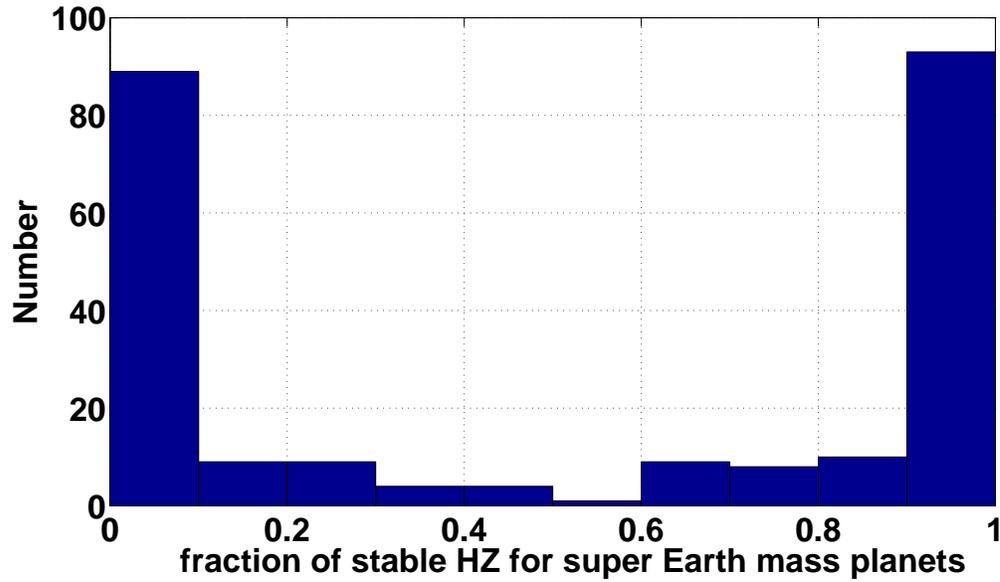}
}
\caption{ Distribution of fraction of stable HZ for hypothetical 
$1 \mearth$ planet (panel a) and $10 \mearth $ planet (panel b), in 
currently observed single planet systems. Of the total systems ($236$)
that we considered, 
nearly $40 \% (95)$ of the systems
have $\ge 90 \%$ of their HZ stable (peak near 1). About $38 \% (91) $
 of the systems
have less than $10 \%$ of their HZ stable (peak near 0). We do not consider
 systems that have planetary masses $> 10 \mjup$
 or if $e > 0.6$.}
\label{fraction}
\end{figure}

For astrobiological purposes, the utility of $\tau_\mathrm{s}$ is 
multi-fold. Not only is it  useful in identifying stable regions within the
HZ of a given system, but it can also provide (based on
 the range of $a$ \& $e$)
what fraction of the HZ is stable.
We have calculated this fraction for all single planet systems in the Exoplanet 
Data Explorer, as of March 25 2010.
 The distribution of fractions of
 currently known 
single planet systems is shown in Fig. \ref{fraction} and tabulated in 
Table \ref{fractable}.  A bimodal distribution can clearly be seen.
Nearly $40 \%$ of the systems have more than $90 \%$ of their HZ stable and $38 \%$ 
of the systems have less than $10 \%$ of their HZ stable. Note that if we include
systems with masses $> 10 \mjup$ and also $e > 0.6$ (which tend to have
$a \sim 1$ AU \citep{wright2009}), the distribution
 will change and there will be relatively fewer HZs that are fully stable.

\section{Summary}
\label{sec5}
We have empirically determined the Lagrange
stability boundary for a planetary system consisting of one terrestrial
mass planet and one massive planet, with initial eccentricities less than 0.6.
Our analysis  shows that for two-planet systems with one terrestrial like
planet and one more massive planet, Eq.(\ref{tauearth}) defines Lagrange stable
configurations and can be used to identify
systems with HZs stable for terrestrial
mass planets. Furthermore, in Table \ref{fractable} we provide
 a catalog of exoplanets,
identifying the fraction of HZ that is Lagrange stable for terrestrial mass 
planets. 
A full version of the table is available in the electronic edition of
the journal\footnote{Updates to this catalog is available at 
\url{http://gravity.psu.edu/~ravi/planets/}.}.

In order to identify stable configurations for a terrestrial planet,
one can calculate a stability boundary (denoted as $\tau_\mathrm{s}$ 
from Eq.(\ref{tauearth}))
 for a given system 
(depending on the  eccentricity and mass 
of the known planet),
and calculate the range of $a$ \& $e$ that can support a terrestrial planet,
as shown in Section 4.
For the transitional region between  unstable to  stable 
($\tau_\mathrm{u} < \bbc < \tau_\mathrm{s}$), a numerical integration should be 
made.
 Our results are in general agreement with previous studies
\citep{MT2003}, \citep{Jones2006} \& \citep{Sandor2007}, but crucially
our approach does not (usually) require a large suite of N-body
 integrations to determine stability.

We have only considered two-planet systems, but  the
 possibility that the star hosts more, currently undetected planets
 is real and
may change the stability boundaries outlined here. However, 
the presence of additional companions will likely
 decrease the size of the stable regions shown in this study.
 Therefore, those systems
that have fully unstable HZs from our analysis will likely continue to have unstable HZs
as more companions are detected (assuming the mass and orbital 
parameters of the
known planet do not change with these additional discoveries). 
 The discovery of an additional planet outside the HZ that
destabilizes the HZ is also an important information.

As more extra-solar planets are discovered, the resources
required to follow-up grows.  Furthermore, as  surveys push to lower 
planet masses, time on large telescopes is required, which is in limited
supply. The study of exoplanets seems poised to transition to an era in which
systems with the potential to host terrestrial mass planets in HZs will be
the focus of surveys.
With limited resources, it will be important to identify systems that can 
actually support a planet in the HZ. The parameter $\tau_{s}$ can 
therefore guide observers as they hunt for the grand prize in exoplanet
research,  an inhabited planet.

Although the current work focuses on  terrestrial mass planets,
the same analysis can be applied to arbitrary configurations that cover
all possible orbital parameters. 
Such a study could represent a significant improvement over the work of
\cite{BG07}. The results presented here show that $\bbc=1$ is not always
the Lagrange stability boundary, as they suggested. An expansion of this 
research to a wider range of planetary and stellar masses and larger
eccentricities could provide an important tool for determining the 
stability and packing of exoplanetary systems.
Moreover, it could reveal an empirical relationship that describes the 
Lagrange stability boundary for two planet systems.
As new planets are discovered in the future, 
the stability maps presented here will guide future research on
 the stability of extra-solar planetary systems.

\acknowledgements

R. K gratefully acknowledges the support of National Science Foundation Grants
No.~PHY 06-53462 and No.~PHY 05-55615, and NASA Grant No.~NNG05GF71G, awarded
to The Pennsylvania State University.  R.B. acknowledges funding
from NASA Astrobiology Institutes's Virtual Planetary Laboratory lead team,
supported by NASA under Cooperative Agreement No. NNH05ZDA001C.
This research has made use of the Exoplanet Orbit Database
and the Exoplanet Data Explorer at exoplanets.org.
The authors acknowledge the Research Computing and Cyberinfrastructure unit of
Information Technology Services at The Pennsylvania State University for
providing HPC resources and services that  contributed to the research
results reported in this paper. URL: http://rcc.its.psu.edu.

%

\clearpage
\thispagestyle{empty}



\clearpage
\thispagestyle{empty}


\clearpage
\thispagestyle{empty}

\clearpage
\thispagestyle{empty}

\clearpage
\thispagestyle{empty}

\clearpage
\begin{deluxetable}{cc}
\tablecaption{Properties of systems considered in this study.}
\tablewidth{0pt}
\tablehead{ \colhead{``Known'' Planet} & \colhead{ a (AU)} }
\startdata 10 M$_\mathrm{jup}$ & (0.25, 4)  \\
\\
 5.6 M$_\mathrm{jup}$ & (0.25, 4)  \\
\\
 3 M$_\mathrm{jup}$ & (0.25, 4)  \\
\\
 1.77 M$_\mathrm{jup}$ & (0.25, 4)  \\
\\
 1 M$_\mathrm{jup}$ & (0.25, 4)  \\
\\
 1.86 M$_\mathrm{Sat}$ & (0.5, 2)  \\
\\
 1 M$_\mathrm{Sat}$ & (0.5, 2)  \\
\\
 56 $\mearth$ & (0.5, 2)  \\
\\
 30 $\mearth$ & (0.5, 2)  \\
\\
 17.7 $\mearth$ & (0.5, 2)  \\
\\
 10 $\mearth$ & (0.5, 2)  \\
 \\
\enddata
\label{table1}
\end{deluxetable}

\clearpage
\begin{table}[h!]
\begin{center}
\begin{threeparttable} 
\caption{ Best fit properties for Eq.\ref{tauearth}.}
\vspace{0.1 in}
\centering
\begin{tabular}{|l|l|l||l|l|}
\hline
\multicolumn{5}{|c|}{~~~~~~~~~~~~~~~~~$1 \mearth$ ~~~~~~~~~~~~~~~~~$ 10 \mearth$} \\
\cline{2-3}
\cline{4-5}
Coefficients & $\tau_\mathrm{s}$ & $\tau_\mathrm{u}$ & $\tau_\mathrm{s}$ & $\tau_\mathrm{u}$\\
\hline
$c_{1}$ & 1.0018 & 1.0098 &0.9868 & 1.0609\\
&&&&\\
$c_{2}$ & -0.0375 & -0.0589 &0.0024 & -0.3547\\
&&&&\\
$c_{3}$ & 0.0633 &  0.04196 &0.1438 & 0.0105\\
&&&&\\
$c_{4}$ & 0.1283 & 0.1078 &0.2155 & 0.6483\\
&&&&\\
$c_{5}$ & -1.0492 & -1.0139 &-1.7093 & -1.2313\\
&&&&\\
$c_{6}$ & -0.2539 & -0.1913 &-0.2485 & -0.0827\\
&&&&\\
$c_{7}$ & -0.0899 & -0.0690 &-0.1827 & -0.4456\\
&&&&\\
$c_{8}$ & -0.0316 & -0.0558 &0.1196 & -0.0279\\
&&&&\\
$c_{9}$ & 0.2349 & 0.1932 &1.8752 & 0.9615\\
&&&&\\
$c_{10}$ & 0.2067 & 0.1577 &-0.0289 & 0.1042\\
&&&&\\
$R^{2}$ & 0.996 & 0.997 &0.931 & 0.977\\
&&&&\\
$\sigma$ & 0.0065 & 0.0061 &0.0257 & 0.0141\\
&&&&\\
Max. dev. & 0.08 & 0.08 &0.15 & 0.05\\
\hline
\end{tabular}
\end{threeparttable}
\end{center}
\label{table0}
\end{table}

\begin{deluxetable}{ccccc}
\tablecaption{Observed parameters of example systems presented in 
\S \ref{sec4}.}
\tablewidth{0pt}
\tablehead{ \colhead{System} &\colhead{M $\sin i$} & \colhead{ a (AU)} & \colhead{e} & \colhead{$M_{\star} (M_\odot)$}}
\startdata Rho CrB & 1.06 $M_\mathrm{jup}$  & 0.23 & 0.06 ($\pm$ 0.028) & 0.97\\
\\
HD 164922 & 0.36 $M_\mathrm{jup}$  & 2.11 & 0.05 ($\pm$ 0.14) & 0.94\\
 \\
GJ 674 & 12 $\mearth$  & 0.039 & 0.20 ($\pm$ 0.02) &  0.35\\
 \\
HD 7924 & 9.26 $\mearth$  & 0.057 & 0.17 ($\pm$ 0.16) &  0.832\\
 \\
\enddata
\label{table2}
\end{deluxetable}

\clearpage
\begin{deluxetable}{cccccccccc}
\tablecaption{Lagrange stable ($\tau_\mathrm{s}$) and unstable ($\tau_\mathrm{u}$) boundaries, 
and the corresponding fraction of habitable zone (FHZ)
stable for terrestrial mass planets in known single planet 
systems. A full version of the table is available in the electronic edition of the {\it Astrophysical Journal.}}
\tablewidth{0pt}
\tablehead{ \colhead{System} &\colhead{$m_\mathrm{1} (M_\mathrm{jup})$} & \colhead{ a (AU)} & \colhead{e} & \colhead{$\tau_\mathrm{s}$} & \colhead{$\tau_\mathrm{u}$} & FHZ & $\tau_\mathrm{s}$ & $\tau_\mathrm{u}$ & FHZ  \\
&&& & $(1 \mearth)$ & $(1 \mearth)$ & $(1 \mearth)$ & $(10 \mearth)$ & $(10 \mearth)$ & $(10 \mearth)$}
\startdata  HD 142b   &     1.3057 &  1.04292 &  0.26 &  0.9347 &  0.9323 &  0.000 &  0.9552 &  0.9320 &  0.000 \\
\\
HD 1237   &    3.3748 &  0.49467 &  0.51 &  0.7407 &  0.7401 &  0.213 &  0.7549 &  0.7450 &  0.000 \\
\\
HD 1461   &    0.0240 &  0.06352 &  0.14 &  0.9920 &  0.9780 &  0.976 &  1.0200 &  0.9200 &  0.959 \\
\\
WASP-1   &    0.9101 &  0.03957 &  0.00 &  1.0022 &  0.9990 &  0.990 &  1.0200 &  0.9980 &  0.991 \\
\\
HIP 2247  &    5.1232 &  1.33884 &  0.54 &  0.7138 &  0.7111 &  0.000 &  0.7490 &  0.7387 &  0.000 \\
\\
\enddata
\label{fractable}
\end{deluxetable}

\end{document}